\pdfoutput=1
\documentclass{aa}

\usepackage{txfonts}
\usepackage{natbib}
\usepackage[colorlinks=true,linkcolor=blue,citecolor=blue,urlcolor=blue]{hyperref}

\begin{document}

\defcitealias{frognerAcceleratedParticleBeams2020}{Paper I}

\title{Accelerated particle beams in a 3D simulation of the quiet Sun}
\subtitle{Effects of advanced beam propagation modelling}
\titlerunning{Effects of advanced beam propagation modelling in 3D}

\author{
    L. Frogner
    \and
    B. V. Gudiksen
}

\institute{
    Institute of Theoretical Astrophysics,
    University of Oslo,
    P.O.Box 1029 Blindern,
    N-0315 Oslo,
    Norway
    \and
    Rosseland Centre for Solar physics (RoCS),
    University of Oslo,
    P.O.Box 1029 Blindern,
    N-0315 Oslo,
    Norway
}

\date{}

\abstract
{
     Charged particles are constantly accelerated to non-thermal energies by the reconnecting magnetic field in the solar atmosphere. Our understanding of the interactions between the accelerated particles and their environment can benefit considerably from three-dimensional atmospheric simulations that account for non-thermal particle beam generation and propagation. In a previous publication, we presented the first results from such a simulation, which considers quiet Sun conditions. However, the original treatment of beam propagation ignores potentially important phenomena like the magnetic gradient forces associated with a converging or diverging magnetic field.
}
{
    Here, we present a more general beam propagation model incorporating magnetic gradient forces, the return current, acceleration by the ambient electric field, corrected collision rates due to the ambient temperature, and collisions with heavier elements than hydrogen and the free electrons they contribute. Neglecting collisional velocity randomisation makes the model sufficiently lightweight to simulate millions of beams. We investigate how each new physical effect in the model changes the non-thermal energy transport in a realistic three-dimensional atmosphere.
}
{
    We applied the method of characteristics to the steady-state continuity equation for electron flux to derive ordinary differential equations for the mean evolution of energy, pitch angle, and flux with distance. For each beam, we solved these numerically for a range of initial energies to obtain the evolving flux spectrum, from which we computed the energy deposited into the ambient plasma.
}
{
    Magnetic gradient forces significantly influence the spatial distribution of deposited beam energy. The magnetic field converges strongly with depth in the corona above loop footpoints. This convergence leads to a small coronal peak in deposited energy followed by a heavy dip caused by the onset of magnetic mirroring. Magnetically reflected electrons carry away 5 to 10\% of the injected beam energy on average. The remaining electrons are relatively energetic and produce a peak in deposited energy below the transition region a few hundred kilometres deeper than they would in a uniform magnetic field. A diverging magnetic field at the beginning of the trajectory, which is common in the simulation, enhances the subsequent impact of magnetic mirroring. The other new physical effects do not qualitatively alter the picture of non-thermal energy transport for the atmospheric conditions under consideration.
}
{}


\maketitle


\section{Introduction}
\label{sec:introduction}

Accelerated particles are one of the less understood outcomes of the magnetic field reconnecting in the solar corona. Magnetic reconnection can bring the particle distribution out of thermal equilibrium and produce beams of non-thermal particles travelling along the magnetic field. The beams can, in some cases, contain a significant portion of the released magnetic energy \citep{lin10100KeV1971, emslieEnergyPartitionTwo2004, emslieGlobalEnergeticsThirtyeight2012}. The effects of accelerated particles are routinely observed in large flares, but their role in smaller energy-release events is unclear.

The interactions between the non-thermal particle beams and their environment have been modelled with ever-increasing levels of sophistication. Early models calculated the mean rate of change in the velocities of the beam particles due to Coulomb collisions with free electrons and protons \citep{brownDirectivityPolarisationThick1972,syrovatskiiHeatingPlasmaHighEnergy1972}. This was later generalised to include collisions with neutral hydrogen, initially for when the fraction of ionised hydrogen is uniform \citep{emslieCollisionalInteractionBeam1978}, subsequently for a non-uniform ionisation fraction \citep{hawleySolarFlareModel1994}. Other studies incorporated the effects of a neutralising return current \citep{knightReverseCurrentSolar1977,emslieEffectReverseCurrents1980} and deflection of particle velocities due to variations in magnetic flux density \citep{leachImpulsivePhaseSolar1981,chandrashekarCollisionalHeatingNonthermal1986}. Instead of calculating mean rates of change, \citet{leachImpulsivePhaseSolar1981} numerically solved the Fokker--Planck equation governing the evolution of a distribution of non-thermal particles. By doing this, they could account for pitch angle diffusion -- the randomisation of directions resulting from the stochastic nature of collisions -- in their beam transport simulations. Further improvements to their model include the incorporation of relativistic effects and energy losses caused by the emission of synchrotron radiation \citep{petrosianDirectivityBremsstrahlungRadiation1985,mctiernanBehaviorBeamsRelativistic1990}. While early treatments of collisions typically ignored the thermal motion of the ambient plasma, corrected expressions accounting for the ambient temperature have later been employed for the mean rate of change in velocity and the rates of energy and pitch angle diffusion \citep{emslieDeterminationTotalInjected2003, gallowayFastElectronSlowingdown2005,jeffreyVariationSolarFlare2014}.

While the models for propagation of non-thermal particle beams have become highly developed, they have mainly been applied to individual beams in isolated one-dimensional (1D) atmospheres, typically combined with hydrodynamics and radiative transfer for the purpose of simulating flares \citep{hawleySolarFlareModel1994,abbettDynamicModelsOptical1999,allredRadiativeHydrodynamicModels2005,liuCOMBINEDMODELINGACCELERATION2009,allredUnifiedComputationalModel2015,allredModelingTransportNonthermal2020,jeffreyRoleEnergyDiffusion2019} or nanoflares \citep{testaEvidenceNonthermalParticles2014,politoInvestigatingResponseLoop2018,bakkeChromosphericEmissionNanoflare2022}. In recent times, three-dimensional (3D) atmospheric simulation codes have been developed that reproduce many of the observed features of both the solar corona and lower atmosphere in a small patch of the Sun outside of active regions, by solving the equations of magnetohydrodynamics (MHD) combined with radiative transfer and thermal conduction \citep{gudiksenStellarAtmosphereSimulation2011,rempelExtensionMURaMRadiative2017}. These simulation codes have not, however, originally accounted for accelerated particles.

In \citet{frognerAcceleratedParticleBeams2020}, hereafter Paper I, we presented a first approach for integrating the modelling of accelerated electrons into a 3D atmospheric simulation. To let the simulated atmosphere drive the spatial distribution and energetics of the non-thermal electron beams, we detected magnetic reconnection sites and applied a simple parametric model for particle acceleration based on local conditions. We then modelled the collective energy transport by millions of beams by tracing beam trajectories from the reconnection sites and computing the energy deposited into the ambient plasma through collisions between non-thermal electrons and ambient particles. To compute the deposited energy, we employed the relatively basic analytical transport model of \citet{emslieCollisionalInteractionBeam1978}, with the extension by \citet{hawleySolarFlareModel1994} to support a non-uniform ionisation fraction.

Here, we present a more realistic model for the propagation of non-thermal electron beams that is still computationally efficient enough to be applied to millions of beams. The model is based on solving the continuity equation for electron flux by transforming it into a set of ordinary differential equations using the method of characteristics. This method has been applied by \citet{craigSolutionElectronContinuity1985, dobranskisExactAnalyticalSolutions2014, dobranskisUpdatedAnalyticalSolutions2015, zharkovaUpdatedAnalyticalSolutions2016} to find analytical solutions to the continuity equation (although see the correction to \citet{dobranskisExactAnalyticalSolutions2014} by \citet{emslieSolutionContinuityEquation2014}). We solve the characteristic equations numerically. By deriving the continuity equation from the Fokker--Planck equation, we show how to incorporate most physical effects accounted for in state-of-the-art models, including magnetic gradient forces, the return current, corrected collisional rates due to the ambient temperature, and contributions to collisions by elements heavier than hydrogen. We additionally consider acceleration along the magnetic field by the ambient electric field, which, to our knowledge, has never been accounted for in models applied to 1D atmospheres. The most notable effects we leave out are collisional randomisation and relativistic effects. We investigate the impact of each new effect on the energy transported by the diverse beams in a 3D atmosphere identical to the one used in \citetalias{frognerAcceleratedParticleBeams2020}.

\section{Methods}
\label{sec:methods}

\subsection{Atmospheric simulation}

We used a snapshot from a 3D simulation of the solar atmosphere to provide a realistic environment for non-thermal particle acceleration and transport. The simulation was performed using the Bifrost code \citep{gudiksenStellarAtmosphereSimulation2011}, which solves the non-ideal equations of MHD, accounting for energy transport through field-aligned thermal conduction and radiative transfer \citep{hayekRadiativeTransferScattering2010, carlssonApproximationsRadiativeCooling2012}. The region of the atmosphere simulated begins at the top of the convection zone, 2.5 Mm below the photosphere, and ends in the corona, 14.3 Mm above the photosphere. It extends 24 Mm in each horizontal direction and uses horizontally periodic boundary conditions. The spatial resolution is 31 km horizontally and between 12 and 80 km vertically, with the finest vertical resolution near the transition region and the coarsest in the upper corona.

The simulation reproduces the basic structure of the quiet solar atmosphere, with a chromosphere and coronal loops heated by magnetic reconnection and acoustic shocks resulting from convective motions below the photosphere. In the snapshot employed for this paper, the atmospheric structure has been further shaped by a magnetic flux emergence event where, 137 minutes of solar time earlier, a 2000 G horizontal magnetic field was injected through the bottom boundary. As it rose to the photosphere, the injected magnetic sheet was broken up by convective motions, enhancing the magnetic field in the network regions between convection cells. The current snapshot, which is the same as the one used in \citetalias{frognerAcceleratedParticleBeams2020}, is characterised by a central bubble of cool plasma (carried upwards by the emerging magnetic field), with increased magnetic reconnection at the upper boundaries of the bubble where the emerging and pre-existing coronal fields meet at a large angle. Further details on this particular simulation can be found in \citet{hansteenEllermanBombsUV2019}. However, we note that the reconnection events in this simulation at the most release energies of order $10^{25}\;\mathrm{erg}$, while most events are much less energetic (see \citetalias{frognerAcceleratedParticleBeams2020}). Thus, we are operating in the energy regime of nanoflares \citep{parkerNanoflaresSolarXray1988}.

\subsection{Accelerated particles}
\label{sec:accelerated-particles}

Magnetic reconnection -- the diffusion of magnetic tension occurring at the interface of domains with oppositely directed magnetic field -- tends to produce environments favourable for the acceleration of ambient charged particles to speeds far exceeding their initial thermal speeds. The strong rotation of the magnetic field induces an electric field in the diffusion region, which, in addition to driving currents in the plasma, may directly accelerate electrons and ions to non-thermal speeds \citep{speiserParticleTrajectoriesModel1965, litvinenkoParticleAccelerationReconnecting1993}. Trapping of charged particles in shrinking magnetic islands \citep{drakeElectronAccelerationContracting2006} and scattering within magnetic turbulence \citep{dmitrukTestParticleAcceleration2003} may respectively lead to first- and second-order Fermi acceleration \citep{fermiGalacticMagneticFields1954, fermiOriginCosmicRadiation1949}. Modelling typically indicates that the resulting energy spectra of accelerated particles are shaped like a power-law.

The particles, confined by the Lorentz force to gyrating trajectories around magnetic field lines, may leave the reconnection sites in coherent beams. In this work, we restrict our attention to accelerated electrons, as these are more readily accelerated to non-thermal speeds than ions due to their low mass. Consequently, they are more likely to transport their energy a significant distance. We can represent a beam of non-thermal electrons in terms of its phase-space distribution function $f(\mathbf{r}, \mathbf{v}, t)$, defined such that $f(\mathbf{r}, \mathbf{v}, t)\;\mathrm{d}^3r\;\mathrm{d}^3v$ is the number of beam electrons within the volume $\mathrm{d}^3r$ around the position $\mathbf{r}$ with velocities within $\mathrm{d}^3v$ of the velocity $\mathbf{v}$ at time $t$. The CGS unit of $f$ is $\mathrm{electrons}/\mathrm{cm}^3/(\mathrm{cm}/\mathrm{s})^3$. We assume that the beam reaches a steady state before the ambient plasma can respond to its presence, so we omit the explicit time dependence. By assuming that the distribution is independent of the azimuthal angle and radial distance of the electrons in their gyrating trajectory around the magnetic field line, we can reduce the phase-space coordinates $(\mathbf{r}, \mathbf{v})$ down to the three independent coordinates $(s, E, \mu)$. Here, $s$ is the distance of the electron from its starting position along the field line. $E = m_\mathrm{e}v^2/2$ is the kinetic energy, where $m_\mathrm{e}$ is the electron mass and $v = \left\lVert\mathbf{v}\right\rVert$ is the total speed. $\mu = \cos(\beta)$ is the cosine of the pitch angle $\beta$ between the velocity and the magnetic field, such that $\mathrm{d}s/\mathrm{d}t = \mu v$. With this choice of coordinates, the volume element is $\mathrm{d}^3r = \mathrm{ds}\;\mathrm{dA}$, where $A$ is the cross-sectional area of the beam, and the velocity volume element is $\mathrm{d}^3v = (v/m_\mathrm{e})\;\mathrm{d}\phi_v\;\mathrm{d}\mu\;\mathrm{d}E$, where $\phi_v$ is the azimuthal angle of $\mathbf{v}$. When using $E$ and $\mu$ as the independent velocity coordinates, it is convenient to use the field-aligned electron flux spectrum, defined by
\begin{equation}
    \label{eq:electron-flux-spectrum-definition}
    F(s, E, \mu)\;\mathrm{d}\mu\;\mathrm{d}E = \int_{\phi_v} \mu vf(s, \mathbf{v})\;\mathrm{d}^3v,
\end{equation}
to represent the beam rather than the phase-space distribution. The quantity in Eq. \eqref{eq:electron-flux-spectrum-definition} is the rate of electrons with energies within $\mathrm{d}E$ of $E$ and pitch angle cosines within $\mathrm{d}\mu$ of $\mu$ flowing through a unit cross-sectional area in the positive magnetic field direction. The CGS unit of $F$ is $\mathrm{electrons}/\mathrm{s}/\mathrm{cm}^2/\mathrm{erg}$.

\subsubsection{Injected distributions}
\label{sec:injected-distributions}

To specify the field-aligned flux spectrum $F_0(E_0, \mu_0) \equiv F(s=0, E, \mu)$ of non-thermal electrons injected into each beam in the simulated atmosphere, we apply the same simple parametric acceleration model as in \citetalias{frognerAcceleratedParticleBeams2020}. We classify a grid cell in the simulation box as a reconnection site if the quantity
\begin{equation}
    \label{eq:krec}
    K = \left\lVert\mathbf{B}\times\left(\nabla\times\left(\left(\frac{\mathbf{E} \cdot \mathbf{B}}{\mathbf{B} \cdot \mathbf{B}}\right)\mathbf{B}\right)\right)\right\rVert,
\end{equation}
exceeds a small threshold. Here, $\mathbf{E}$ and $\mathbf{B}$ are the local macroscopic electric and magnetic fields, respectively. In MHD theory, $K = 0$ is a criterion for the magnetic topology to be conserved \citep{biskampMagneticReconnectionPlasmas2005}, so we detect reconnection by finding where this criterion is sufficiently violated, with $K$ exceeding some small threshold (of the order of $10^{-15}\;\mathrm{G}\;\mathrm{statV}/\mathrm{cm}^2$). We discuss how we chose the threshold value in \citetalias{frognerAcceleratedParticleBeams2020}.

At each reconnection site, we assume that some unspecified time-independent acceleration mechanism maintains a non-thermal power-law tail on top of the local thermal electron distribution. We express this non-thermal component in terms of an injected electron flux spectrum in the form
\begin{equation}
    \label{eq:injected-flux-spectrum}
    F_0(E_0 \geq E_\mathrm{c}, \mu_0) = \mathcal{F}_{\mathrm{beam},0}\frac{\delta - 2}{{E_\mathrm{c}}^2}\left(\frac{E_0}{E_\mathrm{c}}\right)^{-\delta}\delta(\mu_0 - \bar{\mu_0}),
\end{equation}
where $\mathcal{F}_{\mathrm{beam},0}$ is the field-aligned energy flux injected into the beam, $E_\mathrm{c}$ is the lower cut-off energy below which the distribution is empty, and $\delta$ is the power-law index describing how strongly the distribution diminishes with increasing energy. We assume that every injected non-thermal electron has the same initial pitch angle cosine $\mu_0 = \bar{\mu_0}$. This is expressed through the Dirac delta function $\delta(\mu_0 - \bar{\mu_0})$.

To determine the injected energy flux $\mathcal{F}_{\mathrm{beam},0}$, we assume that 20\% of the power released from the magnetic field during reconnection goes into electron acceleration (we discuss the choice of this percentage in \citetalias{frognerAcceleratedParticleBeams2020}). We partition this power between two beams pointed in opposite directions along the magnetic field based on the local alignment of the electric field with the magnetic field. The opposite directions are encoded into the sign that each of the two beams receives for $\bar{\mu_0}$ (and $\mathcal{F}_{\mathrm{beam},0}$). The power assigned to each beam is then converted to an energy flux $\mathcal{F}_{\mathrm{beam},0}$ using the cross-sectional area $A$ of the beam, which is computed based on the extent of the grid cell at the reconnection site. We calculate the lower cut-off energy $E_\mathrm{c}$ as the energy where the resulting non-thermal power-law distribution would intersect the local thermal distribution. The resulting value of $E_\mathrm{c}$ is approximately proportional to the local temperature $T$, with $E_\mathrm{c}$ between 1 and 3 keV at $T = 1\;\mathrm{MK}$. For the power-law index $\delta$, we assume the same fixed value for every beam. Finally, we estimate the magnitude of the initial pitch angle cosine, $|\bar{\mu_0}|$, by assuming that the acceleration process does not alter the initial perpendicular velocity component due to randomised thermal motion. In practice, this results in $|\bar{\mu_0}|$ taking a value slightly below unity for most beams.

\subsubsection{Transport equations}

The evolution of the steady-state field-aligned electron flux spectrum $F(s, E, \mu)$ with distance $s$ is governed by the Fokker--Planck equation:
\begin{multline}
\label{eq:final-fokker-planck}
    \frac{\partial F}{\partial s} + \left(\frac{\mathrm{d}E}{\mathrm{d}s}\right)_\mathrm{!C}\frac{\partial F}{\partial E} + \left(\frac{\mathrm{d}\mu}{\mathrm{d}s}\right)_\mathrm{!C}\frac{\partial F}{\partial \mu} = \left(\frac{1}{E}\left(\frac{\mathrm{d}E}{\mathrm{d}s}\right)_\mathrm{!C} + \frac{1}{\mu}\left(\frac{\mathrm{d}\mu}{\mathrm{d}s}\right)_\mathrm{!C}\right)F \\
     + \frac{\partial}{\partial E}\left(C_E F\right) + \frac{\partial}{\partial \mu}\left(C_\mu F\right) + \frac{\partial^2}{\partial E^2}\left(C_{E^2} F\right) + \frac{\partial^2}{\partial \mu^2}\left(C_{\mu^2} F\right).
\end{multline}
See Appendix \ref{sec:appendix-fokker-planck-equation} for the derivation of this equation from the more general Fokker--Planck equation expressed in terms of $f(\mathbf{r}, \mathbf{v})$. In Eq. \eqref{eq:final-fokker-planck}, $(\mathrm{d}E/\mathrm{d}s)_\mathrm{!C}$ and $(\mathrm{d}\mu/\mathrm{d}s)_\mathrm{!C}$ are the rates of change in $E$ and $\mu$ with distance $s$ due to non-collisional forces. The main such force is the Lorentz force $\mathbf{F}_\mathrm{L}$ on an electron due to the macroscopic electric and magnetic field. In CGS units, it is given as
\begin{equation}
    \label{eq:lorentz-force}
    \mathbf{F}_\mathrm{L} = -e\left(\mathbf{E} + \frac{\mathbf{v}}{c}\times\mathbf{B}\right),
\end{equation}
where $e$ is the elementary charge and $c$ is the speed of light. The electric field $\mathbf{E}$ can accelerate electrons in any direction. We only consider the influence of the component $\mathcal{E}$ of the electric field parallel with the magnetic field direction:
\begin{equation}
    \mathcal{E} = \frac{\mathbf{E} \cdot \mathbf{B}}{B},
\end{equation}
where $B = \left\lVert\mathbf{B}\right\rVert$. We ignore the transverse component of the electric field. A non-zero transverse component causes the electron orbit to drift with velocity $\mathbf{v}_\mathrm{d} = c\mathbf{E} \times \mathbf{B}/B^2$, either breaking azimuthal symmetry or (if the electric field is azimuthally symmetric) gradually changing the gyroradius. In either case, the drift speed will be tiny compared to the parallel speed of the non-thermal electrons. The highest drift speeds in our atmospheric simulation are of order $10^3\;\mathrm{cm}/\mathrm{s}$, far below even the average thermal speed.

We consider two potential contributions to the electric field $\mathbf{E}$. The first is the MHD electric field $\mathbf{E}_\mathrm{fluid}$, which may have a parallel component $\mathcal{E}_\mathrm{fluid}$ when the resistivity is significant (as during magnetic reconnection). The second contribution is the electric field produced in response to the charge displacement that occurs when the accelerated electrons leave the acceleration site. This electric field, which we denote $\mathbf{E}_\mathrm{beam}$, will produce a counter-flowing return current in the background plasma that cancels out the current associated with the beam \citep{knightReverseCurrentSolar1977, emslieEffectReverseCurrents1980}. It is always aligned with the magnetic field, so $\mathbf{E}_\mathrm{beam} = \mathcal{E}_\mathrm{beam} \mathbf{B}/B$. As discussed in \citet{emslieEffectReverseCurrents1980}, we can compute $\mathcal{E}_\mathrm{beam}$ from the beam electron flux. The field-aligned current density associated with the beam at distance $s$ is
\begin{equation}
    J_\mathrm{beam}(s) = -e F_\mathrm{beam}(s),
\end{equation}
where $F_\mathrm{beam}(s)$ is the total field-aligned flux of the beam electrons at distance $s$ (given by Eq. \eqref{eq:field-aligned-electron-flux}). We note that $F_\mathrm{beam}$ is positive if the beam travels along the positive magnetic field direction. Assuming that the return current $J_\mathrm{r}$ exactly neutralises the beam current, we have $J_\mathrm{r}(s) = -J_\mathrm{beam}(s)$. From Ohm's law, the electric field is thus
\begin{equation}
    \label{eq:return-current-electric-field}
    \mathcal{E}_\mathrm{beam}(s) = \eta(s)J_\mathrm{r}(s) = \eta(s)e F_\mathrm{beam}(s),
\end{equation}
where $\eta$ is the electrical resistivity of the ambient plasma parallel to the magnetic field. We compute $\eta$ accounting for electron collisions with protons and neutral hydrogen (see, for example, \citet{chaeElectricResistivityPartially2021}).

The work the total parallel electric field $\mathcal{E} = \mathcal{E}_\mathrm{fluid} + \mathcal{E}_\mathrm{beam}$ does on a beam electron moving with velocity $\mathrm{d}s/\mathrm{d}t = \mu v$ along the magnetic field line is
\begin{equation}
    \label{eq:electric-force-dedt}
    \left(\frac{\mathrm{d}E}{\mathrm{d}t}\right)_\mathrm{E} = -\mu v e\mathcal{E},
\end{equation}
where the subfix E stands for electric. The acceleration along the magnetic field line is
\begin{equation}
    \label{eq:electric-force-parallel-acceleration}
    \left(\frac{\mathrm{d}(\mu v)}{\mathrm{d}t}\right)_\mathrm{E} = -\frac{e\mathcal{E}}{m_\mathrm{e}}.
\end{equation}
The magnetic component of the Lorentz force does no work on the electron since $\mathbf{v} \times \mathbf{B}$ is always perpendicular to the velocity $\mathbf{v}$, so we have
\begin{equation}
    \label{eq:magnetic-force-dedt}
    \left(\frac{\mathrm{d}E}{\mathrm{d}t}\right)_\mathrm{M} = 0,
\end{equation}
where the subfix M stands for magnetic. The only direct effect of the magnetic force is to produce the angular velocity $\mathrm{d}\phi_v/\mathrm{d}t = eB/(m_\mathrm{e}c)$ responsible for the gyrating motion. If the magnetic field is uniform, it does not affect the parallel velocity of the electron. However, if there is a weak gradient in the magnetic field, analysis of the linearised equation of motion shows that the electron will experience a net parallel acceleration. This acceleration, due to the magnetic force averaged over one gyration period, is given by \citep[e.g.][]{bittencourtFundamentalsPlasmaPhysics2004a}
\begin{equation}
    \label{eq:magnetic-gradient-parallel-acceleration}
    \left(\frac{\mathrm{d}(\mu v)}{\mathrm{d}t}\right)_\mathrm{M} = -\frac{M}{m_\mathrm{e}}\frac{\mathrm{d}B}{\mathrm{d}s} = -\left(1 - {\mu}^2\right)\frac{E}{m_\mathrm{e} B}\frac{\mathrm{d}B}{\mathrm{d}s},
\end{equation}
where $M$ is the gyromagnetic moment of the electron. This assumes that the electron experiences no other forces than the magnetic force, which is not strictly true in our model, considering that the electron can be affected by a parallel electric field and collisions. However, any parallel electric field will be weak compared to the magnetic field, and a comparison of the gyrofrequency and collisional frequency for typical atmospheric conditions shows that the electrons will gyrate many times for every collision. Equation \eqref{eq:magnetic-gradient-parallel-acceleration} thus remains a reasonable approximation even in the presence of other forces.

In addition to the Lorentz force, we could also include the reaction force due to the emission of gyromagnetic radiation. This reaction force is given by the Abraham--Lorentz formula \citep[e.g.][]{barutElectrodynamicsClassicalTheory1980a}. However, as discussed by \citet{petrosianDirectivityBremsstrahlungRadiation1985}, who derived expressions for the corresponding rates of change in energy and pitch angle, the influence of the reaction force is significant only for highly relativistic electrons. Hence, there is nothing to gain by accounting for gyromagnetic radiation until the transport model is extended to include relativistic effects. We note that in the simulations performed for this paper, the most relativistic beams have $\delta = 4$ and $E_\mathrm{c} \approx 5\;\mathrm{keV}$. Relativistic electrons with a Lorentz factor exceeding 1.1 account for only 1\% of the energy flux injected into these beams, so disregarding relativistic effects is justified.

Adding Eqs. \eqref{eq:electric-force-dedt} and \eqref{eq:magnetic-force-dedt}, the full rate of change in energy due to non-collisional forces becomes
\begin{equation}
    \label{eq:non-collisional-forces-dedt}
    \left(\frac{\mathrm{d}E}{\mathrm{d}t}\right)_\mathrm{!C} = -\mu v e\mathcal{E},
\end{equation}
while, from Eqs. \eqref{eq:electric-force-parallel-acceleration} and \eqref{eq:magnetic-gradient-parallel-acceleration}, the total parallel acceleration becomes
\begin{equation}
    \label{eq:non-collisional-forces-parallel-acceleration}
    \left(\frac{\mathrm{d}(\mu v)}{\mathrm{d}t}\right)_\mathrm{!C} = -\frac{e\mathcal{E}}{m_\mathrm{e}} - \left(1 - {\mu}^2\right)\frac{E}{m_\mathrm{e} B}\frac{\mathrm{d}B}{\mathrm{d}s}.
\end{equation}
Finally, using $\mathrm{d}s/\mathrm{dt} = \mu v$, expanding the derivative in Eq. \eqref{eq:non-collisional-forces-parallel-acceleration}, applying Eq. \eqref{eq:non-collisional-forces-dedt}, we find
\begin{align}
    \left(\frac{\mathrm{d}E}{\mathrm{d}s}\right)_\mathrm{!C} &= -e\mathcal{E} \label{eq:non-collisional-forces-deds} \\
    \left(\frac{\mathrm{d}\mu}{\mathrm{d}s}\right)_\mathrm{!C} &= -\frac{1 - \mu^2}{2\mu}\left(\frac{e\mathcal{E}}{E} + \frac{\mathrm{d}\ln B}{\mathrm{d}s}\right). \label{eq:non-collisional-forces-dmuds}
\end{align}
Returning to the Fokker--Planck equation, the terms involving the coefficients $C_E$, $C_\mu$, $C_{E^2}$, and $C_{\mu^2}$ in Eq. \eqref{eq:final-fokker-planck} express the evolution of the flux spectrum due to Coulomb collisions between the beam electrons and ambient plasma particles. The coefficients $C_E$ and $C_\mu$ govern the advection of the flux spectrum in energy and pitch angle space, while $C_{E^2}$ and $C_{\mu^2}$ govern the diffusion. General expressions for all the coefficients are derived in Appendix \ref{sec:appendix-fokker-planck-equation}. The results for $C_E$ and $C_\mu$ are
\begin{multline}
    \label{eq:fp-coll-coef-e-general}
    C_E = \frac{2\pi e^4}{\mu E}\left(\sum_\mathrm{c} \left(\frac{m_\mathrm{e}}{m_\mathrm{c}}\mathrm{erf}(u_\mathrm{c}) - \left(1 + \frac{m_\mathrm{e}}{m_\mathrm{c}}\right)u_\mathrm{c}\mathrm{erf}^\prime(u_\mathrm{c})\right){z_\mathrm{c}}^2 n_\mathrm{c}\ln\Lambda_\mathrm{c} \right. \\
    + \left. \sum_\mathrm{N} Z_\mathrm{N}n_\mathrm{N} \ln\Lambda_\mathrm{N}^\prime\right)
\end{multline}
\begin{equation}
    \label{eq:fp-coll-coef-mu-general}
    C_\mu = \frac{\pi e^4\mu}{\mu E^2}\left(\sum_\mathrm{c}\left(\mathrm{erf}(u_\mathrm{c}) - G(u_\mathrm{c})\right){z_\mathrm{c}}^2 n_\mathrm{c} \ln\Lambda_\mathrm{c}  + \sum_\mathrm{N}{Z_\mathrm{N}}^2 n_\mathrm{N} \ln\Lambda_\mathrm{N}^{\prime\prime}\right).
\end{equation}
Each expression has a sum of contributions from collisions with ambient particles of different species. The subscript c denotes a particular species of charged particles, while N denotes a particular species of neutral particles. For each charged particle species c, $m_\mathrm{c}$ is the particle mass, $z_\mathrm{c}$ is the charge number, $n_\mathrm{c}$ is the number density, and $\ln\Lambda_\mathrm{c}$ is the Coulomb logarithm, for which we follow \citet{emslieCollisionalInteractionBeam1978} and use the expression
\begin{equation}
\label{eq:coulomb-logarithm}
    \ln\Lambda_\mathrm{c} = \ln\left(\frac{m_\mathrm{e} m_\mathrm{c}}{m_\mathrm{e} + m_\mathrm{c}}\frac{v^2\eta}{z_\mathrm{c} e^2}\right),
\end{equation}
where $\eta = v/\nu$ is the electron mean free path and $\nu$ is the plasma frequency.  The quantity $u_\mathrm{c} = v/(\sqrt{2}v_\mathrm{tc})$ is the beam electron speed normalised by the thermal speed $v_\mathrm{tc} = \sqrt{k_\mathrm{B}T_\mathrm{c}/m_\mathrm{c}}$, where $k_\mathrm{B}$ is the Boltzmann constant, and $T_\mathrm{c}$ is the temperature of the species c particles. The functions $\mathrm{erf}(u)$ and $\mathrm{erf}^\prime(u)$ are, respectively, the error function and its derivative, while
\begin{equation}
    G(u) = \frac{\mathrm{erf}(u) - u\mathrm{erf}^\prime(u)}{2u^2}.
\end{equation}
For each neutral particle species N, $Z_\mathrm{N}$ is the atomic number, and $n_\mathrm{N}$ is the number density. The effective Coulomb logarithms $\ln\Lambda_\mathrm{N}^\prime$ and $\ln\Lambda_\mathrm{N}^{\prime\prime}$ are associated respectively with friction and velocity diffusion due to collisions with the neutral particles. They are given by \citep{evansAtomicNucleus1955a, snyderMultipleScatteringFast1949}
\begin{align}
    \ln\Lambda_\mathrm{N}^\prime &= \ln\left(\frac{m_\mathrm{e} v^2}{\sqrt{2}I_\mathrm{N}}\right) \label{eq:neutral-coulomb-logarithm-friction} \\
    \ln\Lambda_\mathrm{N}^{\prime\prime} &= \ln\left(\frac{v}{\sqrt{2}{Z_\mathrm{N}}^{1/3} c \alpha}\right), \label{eq:neutral-coulomb-logarithm-diffusion}
\end{align}
where $I_\mathrm{N}$ is the ionisation potential of a species N particle, $c$ is the speed of light, and $\alpha$ is the fine structure constant. We note that while the Coulomb logarithms $\ln\Lambda_\mathrm{c}$, $\ln\Lambda_\mathrm{N}^\prime$, and $\ln\Lambda_\mathrm{N}^{\prime\prime}$ technically vary with beam electron speed $v$, this dependence is relatively slight, so we ignore it and use a representative value corresponding to the mean speed of the initial distribution. Furthermore, the variations due to the specifics of the particular particle species ($m_\mathrm{c}$, $z_\mathrm{c}$, $I_\mathrm{N}$, and $Z_\mathrm{N}$) are even more minor than the variations due to the electron speed. We, therefore, use the electron's value of $\ln\Lambda_\mathrm{c}$ for all charged particle species and the hydrogen atom's value of $\ln\Lambda_\mathrm{N}^\prime$ and $\ln\Lambda_\mathrm{N}^{\prime\prime}$ for all neutral particle species and omit the subscripts.

In this work, we account for collisions with free electrons ($\mathrm{c} = \mathrm{e}$), free protons ($\mathrm{c} = \mathrm{p}$), singly and doubly ionised helium atoms ($\mathrm{c} = \mathrm{HeI}$ and $\mathrm{c} = \mathrm{HeII}$), neutral hydrogen atoms ($\mathrm{N} = \mathrm{NH}$), and neutral helium atoms ($\mathrm{N} = \mathrm{NHe}$). Because the ambient hydrogen and helium atoms have minimal velocities compared to beam electrons, we have $u_\mathrm{c} \gg 1$ for all c except $\mathrm{c} = \mathrm{e}$. We can thus use the asymptotic values $\mathrm{erf}(u) \rightarrow 1$, $\mathrm{erf}^\prime(u) \rightarrow 0$ and $G(u) \rightarrow 0$ for $u \rightarrow \infty$ for collisions with all charged atoms. Since only $u_\mathrm{e}$ remains relevant, we will use the symbol $u$ as a shorthand for $u_\mathrm{e}$. We can neglect the terms in Eq. \eqref{eq:fp-coll-coef-e-general} that include the factor $m_\mathrm{e}/m_\mathrm{c}$ for collisions with the charged atoms since the hydrogen mass $m_\mathrm{H}$ and helium mass $m_\mathrm{He}$ greatly exceed the electron mass. The number densities of the charged atoms can be expressed as $n_\mathrm{p} = x_\mathrm{H}n_\mathrm{H}$, $n_\mathrm{HeI} = x_\mathrm{HeI}n_\mathrm{He}$, and $n_\mathrm{HeII} = x_\mathrm{HeII}n_\mathrm{He}$, where $x_\mathrm{H}$, $x_\mathrm{HeI}$, and $x_\mathrm{HeII}$ are ionisation fractions, and $n_\mathrm{H}$ and $n_\mathrm{He}$ are the total number densities of respectively hydrogen and helium (including both charged and neutral atoms). For the neutral atoms, we can then write the number densities as $n_\mathrm{NH} = (1 - x_\mathrm{H})n_\mathrm{H}$ and $n_\mathrm{NHe} = (1 - x_\mathrm{HeI} - x_\mathrm{HeII})n_\mathrm{He}$. To express all number densities in terms of the hydrogen density $n_\mathrm{H}$, we further define the electron-to-hydrogen ratio $r_\mathrm{e}$ and helium-to-hydrogen ratio $r_\mathrm{He}$ such that the electron density becomes $n_\mathrm{e} = r_\mathrm{e}n_\mathrm{H}$ and the helium density becomes $n_\mathrm{He} = r_\mathrm{He}n_\mathrm{H}$. Equations \eqref{eq:fp-coll-coef-e-general} and \eqref{eq:fp-coll-coef-mu-general} then become
\begin{align}
    C_E &= \frac{2\pi e^4}{\mu E} n_\mathrm{H} \gamma_E(u) \label{eq:fp-coll-coef-e} \\
    C_\mu &= \frac{\pi e^4}{E^2} n_\mathrm{H} \gamma_\mu(u),\label{eq:fp-coll-coef-mu}
\end{align}
where
\begin{multline}
    \label{eq:effective-coulomb-log-e}
    \gamma_E(u) = W_E(u)r_\mathrm{e}\ln\Lambda \\
    + \left(1 - x_\mathrm{H} + 2\left(1 - x_\mathrm{HeI} - x_\mathrm{HeII}\right)r_\mathrm{He}\right)\ln\Lambda^\prime
\end{multline}
and
\begin{multline}
    \label{eq:effective-coulomb-log-mu}
    \gamma_\mu(u) = \left(W_\mu(u)r_\mathrm{e} + x_\mathrm{H} + \left(x_\mathrm{HeI} + 4x_\mathrm{HeII}\right)r_\mathrm{He}\right)\ln\Lambda \\
    + \left(1 - x_\mathrm{H} + 4\left(1 - x_\mathrm{HeI} - x_\mathrm{HeII}\right)r_\mathrm{He}\right)\ln\Lambda^{\prime\prime}
\end{multline}
can be considered effective Coulomb logarithms. The functions
\begin{align}
    W_E(u) &= \mathrm{erf}(u) - 2u\mathrm{erf}^\prime(u) \label{eq:warm-target-contribution-e} \\
    W_\mu(u) &= \mathrm{erf}(u) - G(u) \label{eq:warm-target-contribution-mu}
\end{align}
encapsulate the dependence of the collision coefficients on the ambient temperature.

Equation \eqref{eq:final-fokker-planck} is a linear second-order partial differential equation. It can be solved numerically in many ways, including finite difference methods \citep[e.g.][]{allredModelingTransportNonthermal2020} and stochastic methods \citep[e.g.][]{jeffreyProbingSolarFlare2020}. Unfortunately, these methods are too computationally expensive for simulating huge numbers of beams. To simplify the problem, we ignore the collisional diffusion of the distribution in velocity space by setting the diffusion coefficients $C_{E^2}$ and $C_{\mu^2}$ to zero (we consider the consequences of this approximation in our discussion in Sect. \ref{sec:discussion}). Equation \eqref{eq:final-fokker-planck} can then be written as
\begin{equation}
    \label{eq:continuity-equation}
    \frac{\partial F}{\partial s} + \frac{\partial}{\partial E}\left(\left(\left(\frac{\mathrm{d}E}{\mathrm{d}s}\right)_\mathrm{!C} - C_E\right)F\right) + \frac{\partial}{\partial \mu}\left(\left(\left(\frac{\mathrm{d}\mu}{\mathrm{d}s}\right)_\mathrm{!C} - C_\mu\right)F\right) = S,
\end{equation}
where
\begin{equation}
    \label{eq:continuity-equation-source}
    S = \left(\left(\frac{1}{E} + \frac{\partial}{\partial E}\right)\left(\frac{\mathrm{d}E}{\mathrm{d}s}\right)_\mathrm{!C} + \left(\frac{1}{\mu} + \frac{\partial}{\partial \mu}\right)\left(\frac{\mathrm{d}\mu}{\mathrm{d}s}\right)_\mathrm{!C}\right)F.
\end{equation}
Equation \eqref{eq:continuity-equation} is the continuity equation for the flux spectrum in phase space. The right-hand side, given by Eq. \eqref{eq:continuity-equation-source}, is the source term, which describes the addition or removal of electron flux due to the macroscopic electric field. Without a parallel electric field, the source term evaluates to zero, meaning that the flux spectrum is conserved as it evolves in phase space.

\begin{figure}[!thb]
    \includegraphics{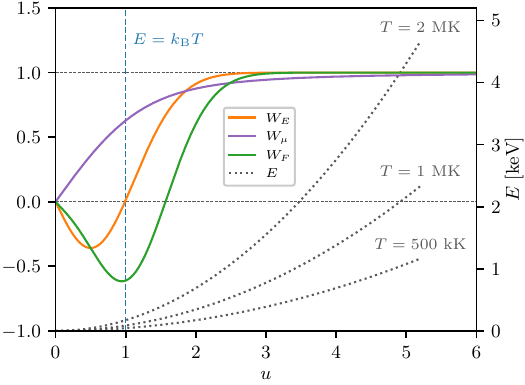}
    \centering
    \caption{Plots of the warm-target contribution functions $W_E$ (orange), $W_\mu$ (purple), and $W_F$ (green) as functions of the normalised speed $u = v/(\sqrt{2}v_\mathrm{t}) = \sqrt{E/k_\mathrm{B}T}$. The dotted grey curves show the mapping between $u$ and energy $E$ for different temperatures $T$. The dashed blue line highlights the location of $u = 1$, corresponding to $E = k_\mathrm{B}T$.}
    \label{fig:warm_target_contribution_functions}
\end{figure}

To solve the continuity equation, we follow \citet{craigSolutionElectronContinuity1985} and transform it into a set of ordinary differential equations using the method of characteristics. The characteristics are curves in the $(s, E, \mu)$ coordinate space that the flux spectrum $F$ follows, given a set of initial conditions. It is convenient to parameterise the curves by the distance $s$. Each curve then describes the evolution of energy $E(s, E_0, \mu_0)$, pitch angle cosine $\mu(s, E_0, \mu_0)$, and flux $F(s, E_0, \mu_0)$ with distance $s$ for a group of beam electrons given their initial values $E_0$, $\mu_0$, and $F_0$ at $s = 0$. The set of ordinary differential equations governing these characteristic curves for Eq. \eqref{eq:continuity-equation} are given by
\begin{align}
    \frac{\mathrm{d}E}{\mathrm{d}s} &= \left(\frac{\mathrm{d}E}{\mathrm{d}s}\right)_\mathrm{!C} - C_E \\
    \frac{\mathrm{d}\mu}{\mathrm{d}s} &= \left(\frac{\mathrm{d}\mu}{\mathrm{d}s}\right)_\mathrm{!C} - C_\mu \\
    \frac{\mathrm{d}F}{\mathrm{d}s} &= \left(\frac{1}{E}\left(\frac{\mathrm{d}E}{\mathrm{d}s}\right)_\mathrm{!C} + \frac{1}{\mu}\left(\frac{\mathrm{d}\mu}{\mathrm{d}s}\right)_\mathrm{!C} + \frac{\partial C_E}{\partial E} + \frac{\partial C_\mu}{\partial \mu}\right)F.
\end{align}
Inserting Eqs. \eqref{eq:non-collisional-forces-deds}, \eqref{eq:non-collisional-forces-dmuds}, \eqref{eq:fp-coll-coef-e}, and \eqref{eq:fp-coll-coef-mu}, this becomes
\begin{align}
    \frac{\mathrm{d}E}{\mathrm{d}s} &= -\frac{2\pi e^4}{\mu E} n_\mathrm{H} \gamma_E(u) - e\mathcal{E} \label{eq:deds} \\
    \frac{\mathrm{d}\mu}{\mathrm{d}s} &= -\frac{\pi e^4}{E^2} n_\mathrm{H} \gamma_\mu(u) - \frac{1 - \mu^2}{2\mu}\left(\frac{e\mathcal{E}}{E} + \frac{\mathrm{d}\ln B}{\mathrm{d}s}\right) \label{eq:dmuds} \\
    \frac{\mathrm{d}F}{\mathrm{d}s} &= -\left(\frac{2\pi e^4}{\mu E^2} n_\mathrm{H} \gamma_F(u) + \frac{1 + \mu^2}{2\mu^2}\frac{e\mathcal{E}}{E} + \frac{1 - \mu^2}{2\mu^2}\frac{\mathrm{d}\ln B}{\mathrm{d}s}\right)F, \label{eq:dfds}
\end{align}
where
\begin{multline}
    \label{eq:effective-coulomb-log-f}
    \gamma_F(u) = W_F(u)r_\mathrm{e}\ln\Lambda \\
    + \left(1 - x_\mathrm{H} + 2\left(1 - x_\mathrm{HeI} - x_\mathrm{HeII}\right)r_\mathrm{He}\right)\ln\Lambda^\prime
\end{multline}
and
\begin{equation}
    \label{eq:warm-target-contribution-f}
    W_F(u) = \mathrm{erf}(u) - \left(2u^2 + \frac{3}{2}\right)u\mathrm{erf}^\prime(u).
\end{equation}
The warm-target contribution functions $W_E$, $W_\mu$, and $W_F$ are plotted in Fig. \ref{fig:warm_target_contribution_functions}. In the cold-target limit, when the beam electron speeds are very high compared to the thermal electron speeds ($u \gg 1$), they all approach unity. When $u$ decreases below $\sim 4$ -- this corresponds to a beam electron energy of 2 to 3 keV for typical coronal temperatures in our simulation -- the functions deviate from unity. As the energy approaches $k_\mathrm{B}T$, $W_E$ and $W_\mu$ decrease, suppressing the collisional rate of energy loss and pitch angle increase. At $E \approx k_\mathrm{B}T$, the beam electrons experience no average loss in energy from collisions with ambient electrons, and for lower energies, they start to gain energy. As evident from the minimum in $W_F$ at $E \approx k_\mathrm{B}T$ (giving a maximally positive contribution to $\mathrm{d}F/\mathrm{d}s$), the influence of the thermal distribution on collisions causes more beam electrons to find themselves with energies close to the mean thermal energy.

\subsubsection{Computing macroscopic quantities}
\label{sec:macroscopic-quantities}

The ultimate objective of determining the electron flux spectrum $F(s, E, \mu)$ is to compute macroscopic quantities associated with the non-thermal electron beam. We can evaluate a macroscopic quantity at a specific distance $s$ by appropriately weighting $F(s, E, \mu)$ and integrating it over $E$ and $\mu$. In general, we can compute the sum $\Sigma(\chi)$ of any individual electron property $\chi(s, E, \mu)$ over all the non-thermal electrons at the distance $s$ by evaluating the following integral:
\begin{equation}
    \Sigma(\chi) = \int_E \int_\mu \chi(s, E, \mu) \frac{F(s, E, \mu)}{\mu v(E)}\;\mathrm{d}\mu\;\mathrm{d}E.
\end{equation}
From Eq. \eqref{eq:electron-flux-spectrum-definition}, we see that this is equivalent to weighting the phase-space distribution function $f$ by $\chi$ and integrating over velocity space. Performing a change of variables from $\mu$ to $\mu_0$, we can write the integral as
\begin{equation}
    \label{eq:general-distribution-integral-mu0-e0}
    \Sigma(\chi) = \int_E \int_{\mu_0} \chi(s, E, \mu_0) \frac{F(s, E, \mu_0)}{\mu(s, E, \mu_0)v(E)} \bigg|\frac{\partial\mu(s, E, \mu_0)}{\partial\mu_0}\bigg|\;\mathrm{d}\mu_0\;\mathrm{d}E.
\end{equation}
By using $(E, \mu_0)$ as primary independent variables instead of $(E, \mu)$ and performing the integral over $E$ and $\mu_0$, we can take advantage of our assumption that every electron in the distribution starts with the same pitch angle cosine $\mu_0 = \bar{\mu_0}$. The Dirac delta function $\delta(\mu_0 - \bar{\mu_0})$ in Eq. \eqref{eq:injected-flux-spectrum} is present in the flux spectrum at all distances $s$. Factoring it out, we can write
\begin{equation}
    F(s, E, \mu_0) = \bar{F}(s, E, \mu_0)\delta(\mu_0 - \bar{\mu_0}).
\end{equation}
Inserting this into Eq. \eqref{eq:general-distribution-integral-mu0-e0} and performing the integral over $\mu_0$, we get
\begin{equation}
    \label{eq:general-distribution-integral-e}
    \Sigma(\chi) = \int_E \chi(s, E, \bar{\mu_0}) \frac{\bar{F}(s, E, \bar{\mu_0})}{\mu(s, E, \bar{\mu_0})v(E)} \bigg|\frac{\partial\mu(s, E, \bar{\mu_0})}{\partial\bar{\mu_0}}\bigg|\;\mathrm{d}E.
\end{equation}
The direction-integrated flux spectrum $\bar{F}(s, E, \bar{\mu_0})$ corresponds to the flux spectrum $F(s, E, \mu_0)$ integrated over $\mu_0$. We have thus reduced the problem of solving for $F$, which is two-dimensional in velocity space, to solving for $\bar{F}$, which is one-dimensional in velocity space, with the additional complication of having to determine the Jacobian determinant $|\partial\mu(s, E, \bar{\mu_0})/\partial\bar{\mu_0}|$.

In terms of the generic integral in Eq. \eqref{eq:general-distribution-integral-e}, the total field-aligned electron flux at the distance $s$ can be written
\begin{equation}
    \label{eq:field-aligned-electron-flux}
    F_\mathrm{beam}(s) = \Sigma(\mu v),
\end{equation}
while the field-aligned energy flux becomes
\begin{equation}
    \label{eq:field-aligned-energy-flux}
    \mathcal{F}_\mathrm{beam}(s) = \Sigma(E \mu v).
\end{equation}
The non-thermal electron number density $n_\mathrm{beam}(s)$ at the distance $s$ is simply
\begin{equation}
    n_\mathrm{beam}(s) = \Sigma(1).
\end{equation}
The corresponding integrand in Eq. \eqref{eq:general-distribution-integral-e} is
\begin{equation}
    \label{eq:dnde}
    \frac{\mathrm{d}n}{\mathrm{d}E} = \frac{\bar{F}(s, E, \bar{\mu_0})}{\mu(s, E, \bar{\mu_0})v(E)} \bigg|\frac{\partial\mu(s, E, \bar{\mu_0})}{\partial\bar{\mu_0}}\bigg|.
\end{equation}
This quantity, the number density of non-thermal electrons per energy interval, provides an informative way of visualising the distribution.

The total power density $Q_\mathrm{beam}(s)$ deposited by the non-thermal electrons at the distance $s$ is found by integrating up the collisional rate of energy loss for the individual electrons:
\begin{equation}
    \label{eq:deposited-power-density}
    Q_\mathrm{beam}(s) = \Sigma\left(-\left(\frac{\mathrm{d}E}{\mathrm{d}t}\right)_\mathrm{coll}\right).
\end{equation}
The collisional rate of change in energy is found by taking the collisional part of Eq. \eqref{eq:deds} and multiplying with $\mathrm{d}s/\mathrm{d}t = \mu v$:
\begin{equation}
    \label{eq:dedt-coll}
    \left(\frac{\mathrm{d}E}{\mathrm{d}t}\right)_\mathrm{coll} = -\frac{2\pi e^4v}{E} n_\mathrm{H} \gamma_E(u).
\end{equation}
We find the total heating power per volume $Q(s)$ resulting from the beam by adding the contribution $Q_\mathrm{r}(s)$ from the resistive heating due to the return current $J_\mathrm{r}$:
\begin{equation}
    \label{eq:total-heating-power-density}
    Q(s) = Q_\mathrm{beam}(s) + Q_\mathrm{r}(s),
\end{equation}
where
\begin{equation}
    \label{eq:return-current-resistive-heating}
    Q_\mathrm{r}(s) = \eta(s) {J_\mathrm{r}(s)}^2 = e^2 \eta(s) {F_\mathrm{beam}(s)}^2.
\end{equation}

\subsubsection{Numerical solution}
\label{sec:numerical-solution}

For each non-thermal electron beam, we obtain its trajectory by tracing the magnetic field line from the reconnection site in the appropriate direction (as determined by the sign of $\bar{\mu_0}$) in the same manner as in \citetalias{frognerAcceleratedParticleBeams2020}. In \citet{frognerImplementingAcceleratedParticle2022}, we covered the procedure for accurately tracing the magnetic field lines in detail. In tandem with tracing the trajectory, we integrate Eqs. \eqref{eq:deds}, \eqref{eq:dmuds}, and \eqref{eq:dfds} simultaneously to obtain the evolution of energy $E(s, E_0, \bar{\mu_0})$, pitch angle cosine $\mu(s, E_0, \bar{\mu_0})$ and flux $\bar{F}(s, E_0, \bar{\mu_0})$ with distance $s$.

We start with a set of initial energies $E_0^{(i)}$ and pitch angle cosines $\mu_0^{(i)}$ for $i = 0 \ldots (N-1)$ at $s = 0$. The energies are evenly distributed in log space with a spacing $\Delta\log_{10}E$, and the pitch angle cosines all have the same value $\bar{\mu_0}$. We then evaluate $\bar{F}_0(E_0, \bar{\mu_0})$ (Eq. \eqref{eq:injected-flux-spectrum} without the Dirac delta function) to obtain a corresponding set of injected flux values, $\bar{F}_0^{(i)} = \bar{F}_0(E_0^{(i)}, \bar{\mu_0})$. Using Eqs. \eqref{eq:deds}, \eqref{eq:dmuds}, and \eqref{eq:dfds}, we advance $E_0^{(i)}$, $\mu_0^{(i)}$, and $\bar{F}_0^{(i)}$ a small distance $\Delta s$ (with the same sign as $\bar{\mu_0}$) using a third-order Runge--Kutta method, to obtain their values $\tilde{E_1}^{(i)}$, $\tilde{\mu_1}^{(i)}$, and $\tilde{F_1}^{(i)}$ at $s_1 = \Delta s$. After the step, some of the lowest energies may have reached zero, meaning that the corresponding particles have been thermalised. If we kept advancing the remaining particles in the same manner, we would soon be left with too few non-zero energies to represent the distribution properly. Moreover, because the derivatives are larger in magnitude for smaller $E$ and $\mu$, the spacing between the lower energies would increase rapidly and amplify the undersampling problem.

To avoid this, we apply a re-meshing procedure after each step. We calculate a new set of energies $E_1^{(i)} = 10^{m\Delta\log_{10}E} E_0^{(i)}$ for $i = 0 \ldots (N-1)$, where $m$ is an integer that may be positive or negative. The new energies are equal to the previous energies $E_0^{(i)}$ except shifted up or down in log space by a whole number $m$ of the interval $\Delta\log_{10}E$. We set $m$ so that the first re-meshed energy $E_1^{(0)}$ is as small as possible while still exceeding the smallest non-zero advanced energy $\tilde{E_1}^{(i)}$. In this way, we ensure good sampling coverage even if the distribution shifts significantly in energy. We then calculate the piecewise linear functions $\tilde{\mu}(E)$ and $\tilde{F}(E)$ that yield $\tilde{\mu}(\tilde{E_1}^{(i)}) = \tilde{\mu_1}^{(i)}$ and $\tilde{F}(\tilde{E_1}^{(i)}) = \tilde{F_1}^{(i)}$ for all non-zero $\tilde{E_1}^{(i)}$. For all re-meshed energies $E_1^{(i)}$ not exceeding the highest advanced energy $\tilde{E_1}^{(n-1)}$, we use these interpolating functions to calculate $\mu_1^{(i)} = \tilde{\mu}(E_1^{(i)})$ and $\bar{F}_1^{(i)} = \tilde{F}(E_1^{(i)})$.

For the re-meshed energies exceeding $\tilde{E_1}^{(N-1)}$, we have no samples of the flux spectrum and thus can no longer rely on interpolation. Instead, by selecting a sufficiently high upper limit for the initial energies, we can ensure that the highest advanced energy $\tilde{E_1}^{(N-1)}$ is large enough that the following high-energy limit versions of Eqs. \eqref{eq:deds}, \eqref{eq:dmuds}, and \eqref{eq:dfds} are valid:
\begin{align}
    \left(\frac{\mathrm{d}E}{\mathrm{d}s}\right)_\mathrm{high} &= - e\mathcal{E} \label{eq:deds-hel}  \\
    \left(\frac{\mathrm{d}\mu}{\mathrm{d}s}\right)_\mathrm{high} &= -\frac{1 - \mu^2}{2\mu} \frac{\mathrm{d}\ln B}{\mathrm{d}s} \label{eq:dmuds-hel} \\
    \left(\frac{\mathrm{d}F}{\mathrm{d}s}\right)_\mathrm{high} &= -\frac{1 - \mu^2}{2\mu^2}\frac{\mathrm{d}\ln B}{\mathrm{d}s}F. \label{eq:dfds-hel}
\end{align}
In the high-energy limit, the influence of collisions is negligible. Only the field-aligned electric field affects the energy of a particle, and only the magnetic gradient force affects the pitch angle and flux. From Eq. \eqref{eq:deds-hel}, the initial energy of a high-energy electron with energy $E$ at the distance $s$ is given by
\begin{equation}
    \label{eq:e-hel}
    E_{\mathrm{high},0}(s, E) = E + \int_0^s e\mathcal{E}(s^\prime)\;\mathrm{d}s^\prime.
\end{equation}
Integrating Eq. \eqref{eq:dmuds-hel}, the pitch angle cosine at the distance $s$ for a high-energy electron with initial pitch angle cosine $\mu_0$ becomes
\begin{equation}
    \label{eq:mu-hel}
    \mu_\mathrm{high}(s, \mu_0) = \mathrm{sgn}\left(\mu_0\right)\sqrt{1 - (1 - {\mu_0}^2)\exp\left(\int_0^s \frac{\mathrm{d}\ln B}{\mathrm{d}s} (s^\prime)\;\mathrm{d}s^\prime\right)}.
\end{equation}
Combining Eqs. \eqref{eq:dmuds-hel} and \eqref{eq:dfds-hel} and integrating, we obtain the expression for the flux of a high-energy electron at the distance $s$ given an initial flux $F_0$:
\begin{equation}
    \label{eq:f-hel}
    F_\mathrm{high}(s, \mu_0, F_0) = \frac{\mu_\mathrm{high}(s, \mu_0)}{\mu_0}F_0.
\end{equation}
For all $E_1^{(i)}$ above the highest advanced energy $\tilde{E_1}^{(N-1)}$, we can obtain the value of $\mu_1^{(i)}$ from Eq. \eqref{eq:mu-hel}. To calculate $\bar{F}_1^{(i)}$, we evaluate Eq. \eqref{eq:e-hel} to find the initial energy of the electron, sample the initial flux spectrum $\bar{F}_0$ at this energy and insert the resulting initial flux into Eq. \eqref{eq:f-hel}:
\begin{align}
    \mu_1^{(i)} &= \mu_\mathrm{high}(s_1, \bar{\mu_0}) \\
    \bar{F}_1^{(i)} &= \frac{\mu_1^{(i)}}{\bar{\mu_0}}\bar{F}_0(E_{\mathrm{high},0}(s_1, E_1^{(i)}), \bar{\mu_0}).
\end{align}
When the re-meshing is complete, we advance $E_1^{(i)}$, $\mu_1^{(i)}$, and $\bar{F}_1^{(i)}$ to obtain $\tilde{E_2}^{(i)}$, $\tilde{\mu_2}^{(i)}$, and $\tilde{F_2}^{(i)}$ at $s_2 = 2\Delta s$, re-mesh these to $E_2^{(i)}$, $\mu_2^{(i)}$, and $\bar{F}_2^{(i)}$, and repeat the procedure. We update the values of the integrals in Eqs. \eqref{eq:e-hel} and \eqref{eq:mu-hel} continually during propagation. With the sets of values $E_k^{(i)}$ and $\bar{F}_k^{(i)}$, we then have a discrete version of the flux spectrum $\bar{F}(s, E, \bar{\mu_0})$ at $s_k = k\Delta s$.

In addition to the flux spectrum $\bar{F}(s, E, \bar{\mu_0})$, we need to determine the Jacobian determinant $|\partial\mu(s, E, \bar{\mu_0})/\partial\bar{\mu_0}|$. It can be calculated analytically for simplified versions of Eqs. \eqref{eq:deds} and \eqref{eq:dmuds} -- ignoring ionisation fraction variations, magnetic gradient forces, electric fields and the ambient temperature -- by solving for $E(s, E_0, \mu_0)$ and $\mu(s, E_0, \mu_0)$, combining these to obtain $\mu(s, E, \mu_0)$ and computing its partial derivative with respect to $\mu_0$. Unfortunately, this is not possible for the more general case. Instead, we compute the Jacobian numerically by evolving a second set of electrons with the same initial energies but with the initial pitch angle cosine $\bar{\mu_0}^\prime = (1 - \varepsilon)\bar{\mu_0}$ reduced by a small fraction $\varepsilon$\footnote{We use $\varepsilon = 10^{-8}$, but this can be varied significantly with equivalent results. For the simplified case when the analytical Jacobian determinant is known, the computed Jacobian is a reasonable match with the exception of some noisy variation at the lowest energies. This variation does not significantly affect the integrated result, however.}. So at any given depth $s_k$, we will have the pitch angle cosines $\mu_k^{\prime(i)}$ for the "perturbed" electrons with $\mu_0^{(i)} = \bar{\mu_0}^\prime$ in addition to the original solution values $E_k^{(i)}$, $\mu_k^{(i)}$, and $E_{0,k}^{(i)}$ for $\mu_0^{(i)} = \bar{\mu_0}$. We note that we advance the perturbed electrons in tandem with those in the original distribution and use the same value for $m$ when re-meshing them. Consequently, after re-meshing, the perturbed electrons will always have the same energies $E_k^{(i)}$ as the original electrons. We then compute the Jacobian determinant as follows:
\begin{equation}
    \bigg|\frac{\partial \mu}{\partial \bar{\mu_0}}\bigg|_k^{(i)} = \bigg|\frac{\mu_k^{(i)} - \mu_k^{\prime(i)}}{\bar{\mu_0} - \bar{\mu_0}^\prime}\bigg|.
\end{equation}
At each distance $s_k$, we use the evolved electron properties $E_k^{(i)}$, $\mu_k^{(i)}$, and $E_{0,k}^{(i)}$ and flux spectrum $\bar{F}_k^{(i)}$ along with the Jacobian determinant $|\partial\mu/\partial\bar{\mu_0}|_k^{(i)}$ to numerically evaluate Eq. \eqref{eq:field-aligned-electron-flux} for the total field-aligned electron flux $F_\mathrm{beam}(s_k)$ and Eq. \eqref{eq:total-heating-power-density} for the heating power density $Q(s_k)$. We use $F_\mathrm{beam}(s_k)$ to calculate the return current resistive heating $Q_\mathrm{r}(s_k)$ (Eq. \eqref{eq:return-current-resistive-heating}) and to estimate the return current electric field $\mathcal{E}_\mathrm{beam}(s_{k+1})$ (Eq. \eqref{eq:return-current-electric-field}) for the following distance $s_{k+1}$. We also use the total flux to decide when the beam has lost enough energy that we can terminate propagation.

\section{Results}
\label{sec:results}

The transport model presented in this paper, hereafter referred to as the continuity equation characteristics (CEC) model, can account for a variety of additional physical effects that are ignored in the analytical transport model employed in \citetalias{frognerAcceleratedParticleBeams2020} (originally from \citet{emslieCollisionalInteractionBeam1978} and \citet{hawleySolarFlareModel1994}). While significantly more computationally intensive than the analytical model, the CEC model remains sufficiently lightweight to be applied on the scale of millions of beams. We can thus compare it directly with the analytical model in the same atmospheric simulation. Our main objective with the results presented in this paper is to highlight the changes in the spatial distribution of beam energy deposition $Q$ resulting from accounting for the additional physical effects.

To verify the CEC model, we ran it under the same assumptions as the analytical model\footnote{Technically, the models can not be made completely equivalent when the ionisation fraction is non-uniform because of an approximation made by \citet{hawleySolarFlareModel1994} in deriving the analytical expression for $Q(s)$. This discrepancy does not prevent a good match in practice, though.} and compared the resulting energy deposition. To match the CEC model with the analytical model, we neglected collisions with helium by setting the helium-to-hydrogen ratio $r_\mathrm{He}$ to zero, we ignored the ambient temperature by setting $W_E$, $W_\mu$, and $W_F$ to unity, we omitted electric fields and magnetic gradient forces by setting $\mathcal{E}$ and $\mathrm{d}\ln B/\mathrm{d}s$ to zero, and we left out ambient electrons from other elements than hydrogen by setting the electron-to-hydrogen ratio $r_\mathrm{e}$ equal to the hydrogen ionisation fraction $x_\mathrm{H}$. The results showed good agreement between the analytical and CEC models, indicating that the latter is sound.

In each of the following sections, we present the results from running the CEC model with one of the additional physical effects included and the others ignored. All runs use $N = 300$ energies to represent the electron flux spectrum $\bar{F}$ and a power-law index of $\delta = 4$ for the injected flux spectrum. To avoid potential confusion, we do not distinguish whether beams travel along the positive or negative magnetic field direction in the results. Hence, $\mu$ and $s$ can always be considered positive.

\subsection{With magnetic gradient forces}

\begin{figure}[!thb]
    \includegraphics{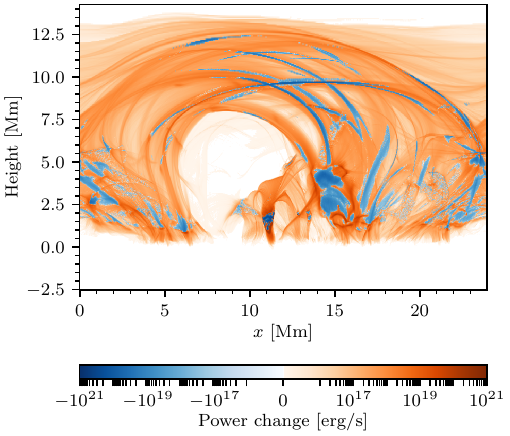}
    \centering
    \caption{Net change in heating power due to non-thermal electrons in the atmospheric simulation accumulated over the $y$-axis of the simulation box. Approximately 1.25 million beams were simulated using the Fokker--Planck-based model, with magnetic gradient forces accounted for in the electron transport equations. Blue regions have a net reduction in heating power resulting from released reconnection energy that would otherwise be released as local heating instead of going into particle acceleration. Orange regions indicate where the non-thermal energy is deposited as heat.}
    \label{fig:xz_power_change_beams_characteristics_4_magmir}
\end{figure}

The run that included magnetic gradient forces, where the $\mathrm{d}\ln B/\mathrm{d}s$ factor in Eq. \eqref{eq:dmuds} was allowed to be non-zero, revealed some significant changes to the energy deposition in the atmospheric simulation compared to the results from the analytical model. Figure \ref{fig:xz_power_change_beams_characteristics_4_magmir} shows the net beam heating power accumulated over the $y$-axis of the simulation box for the run with magnetic gradient forces. This figure is analogous to Fig. 10 in \citetalias{frognerAcceleratedParticleBeams2020}, which shows the corresponding result produced with the analytical model. In both cases, regions of particle acceleration, indicated by negative power (blue colour) due to the absorption of reconnection energy by new beams, appear along the interfaces of misaligned coronal loops and near loop footpoints directly above the transition region (TR). No difference in the distribution of acceleration regions is to be expected between the models, as the treatment of particle acceleration is identical. The distribution of deposited energy (orange colour) is broadly similar for both models, with the most intense beam heating occurring close to the boundaries of the acceleration regions and in the TR. However, for the model with magnetic gradient forces, the energy deposition in the corona is somewhat higher. At the same time, the beams do not penetrate as deep into the chromosphere, stopping abruptly around 500 km above the photosphere.

\begin{figure}[!thb]
    \includegraphics{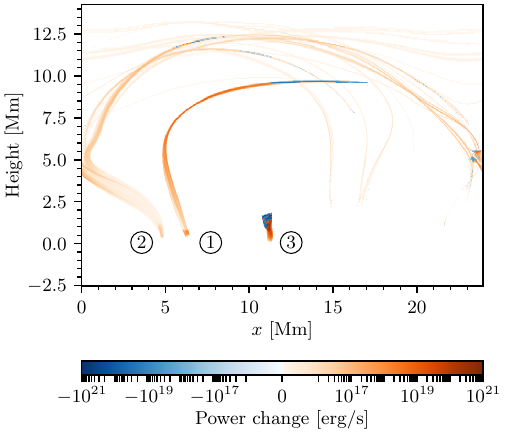}
    \centering
    \caption{Three selected sets of electron beams, plotted in the same manner as in Fig. \ref{fig:xz_power_change_beams_characteristics_4_magmir}. Sets 1, 2, and 3 consist of approximately 4700, 1300, and 2800 beams, respectively.}
    \label{fig:xz_power_change_selected_beams_characteristics_4_magmir}
\end{figure}

To enable a closer analysis of the differences made by magnetic gradient forces, we extracted three separate sets of beams demonstrating different energy transport scenarios. Figure \ref{fig:xz_power_change_selected_beams_characteristics_4_magmir} displays the three sets of beams. The figure is akin to Fig. \ref{fig:xz_power_change_beams_characteristics_4_magmir}, but contains only the net beam heating power due to the selected sets of beams. In set 1, the electrons are accelerated at the top of a coronal loop and follow one of the loop legs down into the TR. In set 2, acceleration occurs at various locations in the corona, but the resulting bundles of beams converge at the same location in the TR. In set 3, the acceleration region is a strong current sheet situated just above the TR, and the non-thermal electrons are ejected downwards in a coherent bundle. These sets correspond to the ones selected for the results in \citetalias{frognerAcceleratedParticleBeams2020}. In the following, we analyse the energy deposition for representative beams in the three sets.

\subsubsection{Coronal loop (set 1)}

\begin{figure}[!thb]
    \includegraphics{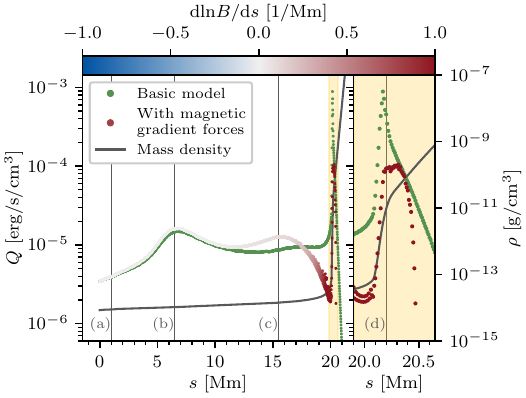}
    \centering
    \caption{Deposited power density $Q$ as a function of propagation distance for a representative beam in set 1 in Fig. \ref{fig:xz_power_change_selected_beams_characteristics_4_magmir} for two separate simulations, one with magnetic gradient forces (points with colour varying between blue and red) and one without (green points). The colours for the former simulation indicate the local relative change in magnetic flux density $B$ with distance $s$. Blue colours imply that the magnetic field decreases in strength in the direction of propagation, while red colours imply that the magnetic field increases in strength. The local plasma mass density is shown as a solid grey curve. The right panel with a yellow background gives a magnified view of the plot in the left panel around the lower atmosphere, over the distance range indicated by the yellow band in the left panel. The labelled vertical lines mark locations of interest along the trajectory: (a) is near the site of injection, (b) is at the first coronal peak in $Q$, (c) is at the second coronal peak present only in the simulation with magnetic gradient forces, and (d) is near the peaks in $Q$ below the transition region. The beam has $\mathcal{F}_{\mathrm{beam},0} \approx 1.4 \cdot 10^4\;\mathrm{erg}/\mathrm{s}/\mathrm{cm}^2$ and $E_\mathrm{c} \approx 5\;\mathrm{keV}$.}
    \label{fig:heating_profile_bundle_1_beam_6265_magmir}
\end{figure}

Due to the high coherency of the beams in set 1, they all have a very similar heating profile, so it suffices to look at one of them. A comparison of $Q(s)$ between the runs with and without magnetic gradient forces for one of the beams is shown in Fig. \ref{fig:heating_profile_bundle_1_beam_6265_magmir}. In the basic model, $Q$ has a peak in the corona around $s = 6.5\;\mathrm{Mm}$ and another, much stronger peak at $s = 20.15\;\mathrm{Mm}$ in the TR. The location of the TR and chromosphere can be inferred from the mass density profile in the figure. When magnetic gradient forces are included, a second coronal peak appears around $s = 15.5\;\mathrm{Mm}$, followed by a steep drop in $Q$ and then a peak in the TR and chromosphere that is smaller but approximately 100 km deeper than the peak in the basic model.

\begin{figure}[!thb]
    \includegraphics{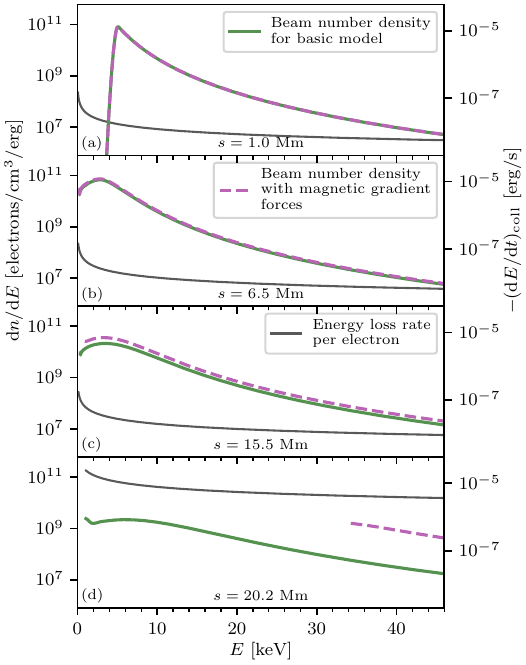}
    \centering
    \caption{The distribution $\mathrm{d}n/\mathrm{d}E$ of beam electrons over energy $E$ at the four separate distances $s$ (with a separate panel for each distance) indicated in Fig. \ref{fig:heating_profile_bundle_1_beam_6265_magmir}, both for a simulation that ignores (solid green curve) and includes (dashed purple curve) magnetic gradient forces. Also included is the collisional rate of energy loss for a single electron, $-(\mathrm{d}E/\mathrm{d}t)_\mathrm{coll}$, from Eq. \eqref{eq:dedt-coll} (solid grey curve). Summing a distribution curve and energy loss rate curve would yield a log-space curve of the integrand for the deposited power density $Q_\mathrm{beam}$ (Eq. \eqref{eq:deposited-power-density}).}
    \label{fig:distributions_bundle_1_beam_6265}
\end{figure}

It is informative to consider the microscopic physical mechanisms leading to these features in the heating profile. To aid the discussion, Fig. \ref{fig:distributions_bundle_1_beam_6265} displays the non-thermal electron number distribution $\mathrm{d}n/\mathrm{d}E$ as a function of energy $E$ for the two models at some selected distances. The injected distribution has the most electrons at the lower cut-off energy $E_\mathrm{c} \approx 5\;\mathrm{keV}$, about 100 times more than at $15\;\mathrm{keV}$ (see panel (a) in Fig. \ref{fig:distributions_bundle_1_beam_6265}, where the distributions are very close to the injected ones). As the electrons begin to propagate, the least energetic electrons lose energy and parallel velocity rapidly, their rate of loss with distance increasing as their energy decreases (due to the $1/E$ factor in Eq. \eqref{eq:deds} and $1/E^2$ factor in Eq. \eqref{eq:dmuds}) in a runaway deceleration. This produces a steady rise in $Q$ with distance, peaking at the point where the abundant least energetic electrons have lost all their excess energy and joined the thermal distribution (panel (b) in Fig. \ref{fig:distributions_bundle_1_beam_6265}).

The initially more energetic electrons experience a slower rate of energy and parallel velocity loss at first (contributing a small part to $Q$) but eventually slow down enough to go through the same rapid deceleration that leads to thermalisation. Because there are fewer of these more energetic electrons (the peak number densities decrease from panel (b) to (c) in Fig. \ref{fig:distributions_bundle_1_beam_6265}), the total power deposited as they become thermal is smaller than that deposited when the least energetic electrons were thermalised, so $Q$ now decreases with distance.

When the electrons enter a region of denser plasma, as in panel (d) in Fig. \ref{fig:distributions_bundle_1_beam_6265}, the collision rate increases, leading to more rapid energy loss and hence an instant proportional increase in $Q$ (due to the $n_\mathrm{H}$ factor in Eq. \eqref{eq:dedt-coll}). The increased collision rate is evident when comparing the rate of energy loss $(\mathrm{d}E/\mathrm{d}t)_\mathrm{coll}$ in panel (d) to the other panels. But with the higher rate of collisions, thermalisation depletes the distribution from the lowest-energy end more rapidly, causing a steeper decrease in $Q$ with distance. This is the reason for the falloff of $Q$ following the TR peak for the basic model in Fig. \ref{fig:heating_profile_bundle_1_beam_6265_magmir}. Entering a region of lower plasma density would have the opposite effect, with less energy deposition and slower depletion of the distribution.

When the electrons experience a rising magnetic flux density as they propagate, the magnetic gradient force deflects the electron velocities, transforming some parallel velocity into transverse velocity and hence reducing their pitch angle cosine $\mu$ (a positive $\mathrm{d}\ln B/\mathrm{d}s$ gives a proportionally negative contribution to $\mathrm{d}\mu/\mathrm{d}s$ in Eq. \eqref{eq:dmuds}). A reduction in $\mu$ means less of the electron's velocity is directed forwards, so it must endure more collisions to advance along the field line. This leads to an increased rate of energy loss with distance (hence the $1/\mu$ factor in Eq. \eqref{eq:deds}).

Less energetic electrons tend to be more susceptible to magnetic deflection because their velocities typically are partially deflected already due to their more frequent collisions. As Eq. \eqref{eq:dmuds} shows, the magnetic contribution to $\mathrm{d}\mu/\mathrm{d}s$ increases with decreasing $\mu$. Still, because the deflection rate is not directly dependent on electron energy, arbitrarily energetic electrons in the distribution can be affected if their velocities are not entirely parallel to the magnetic field. And once magnetic deflection begins, it proceeds equally rapidly regardless of energy. As a result, the magnetic gradient force causes even the more energetic electrons in the distribution to lose energy faster with distance, making them join the less energetic part of the distribution sooner. This positive contribution to the less energetic part of the distribution (evident from comparing the two models in panel (c) in Fig. \ref{fig:distributions_bundle_1_beam_6265}), where most of the energy transfer to the ambient plasma takes place, leads to an increase in $Q$. This is the reason for the second coronal peak in $Q$ for the model with magnetic gradient forces in Fig. \ref{fig:heating_profile_bundle_1_beam_6265_magmir}.

The increase in $Q$ is only temporary, as it is counteracted by the loss of electrons whose velocities become entirely transverse, $\mu = 0$, and thus are left behind. This initially only happens for electrons whose energies are nearly lost already (as in panel (c) in Fig. \ref{fig:distributions_bundle_1_beam_6265}, where the distribution affected by magnetic gradient forces vanishes below $E \approx 1\;\mathrm{keV}$), but subsequently for electrons with more and more energy remaining (the distribution vanishes at increasingly large energies between panel (c) and (d)). The eventual result of the loss of increasingly energetic electrons is a steep drop in $Q$ with distance, as seen after the second coronal peak for the model with magnetic gradient forces in Fig. \ref{fig:heating_profile_bundle_1_beam_6265_magmir}.

After leaving behind the forward-propagating beam, the lost electrons keep accelerating in the opposite direction due to the magnetic gradient force and begin travelling back along the field line in the direction they came from. This phenomenon, known as magnetic mirroring, is not accounted for in the CEC model because the transport equations (Eqs. \eqref{eq:deds}, \eqref{eq:dmuds}, and \eqref{eq:dfds}), being parameterised by distance $s$, become singular for $\mu(s) = 0$. In the model, the energy contained in reflected electrons is lost. For the beam we are considering, the fraction of the injected flux lost to reflected electrons is only a few percent.

An implication of the gradual reflection of increasingly energetic electrons caused by magnetic gradient forces can be seen when the distribution enters a region dense enough to absorb its remaining energy. Because the electrons remaining are relatively energetic (evident from the very high average energy in the distribution affected by magnetic gradient forces compared to the unaffected one in panel (d) in Fig. \ref{fig:distributions_bundle_1_beam_6265}), they penetrate deeper into the region before becoming thermal. Consequently, the peak of deposited power occurs deeper than it would if the less energetic electrons filtered out by magnetic gradient forces remained in the distribution. This explains the deeper peak location in $Q$ for the model with magnetic gradient forces in the right panel of Fig. \ref{fig:heating_profile_bundle_1_beam_6265_magmir}.

\subsubsection{Converging bundles (set 2)}
\label{sec:converging-bundles}

\begin{figure}[!thb]
    \includegraphics{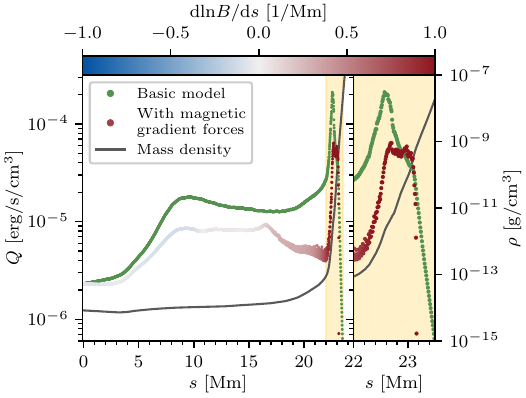}
    \centering
    \caption{Like Fig. \ref{fig:heating_profile_bundle_1_beam_6265_magmir}, but for an electron beam in set 2 starting at $x = 7.1\;\mathrm{Mm}$ and a height of 12.1 Mm in Fig. \ref{fig:xz_power_change_selected_beams_characteristics_4_magmir}. The beam has $\mathcal{F}_{\mathrm{beam},0} \approx 1.1 \cdot 10^4\;\mathrm{erg}/\mathrm{s}/\mathrm{cm}^2$ and $E_\mathrm{c} \approx 5.4\;\mathrm{keV}$.}
    \label{fig:heating_profile_bundle_2_beam_758_magmir}
\end{figure}

Most of the beams in set 2 -- including the majority of beams accelerated near the height of 5 Mm to the far right in Fig. \ref{fig:xz_power_change_selected_beams_characteristics_4_magmir} -- follow a field line downwards in a converging magnetic field and thus have a heating profile qualitatively similar to that in Fig. \ref{fig:heating_profile_bundle_1_beam_6265_magmir}. Some beams, however, experience a diverging magnetic field (that is, a reduction in the magnetic flux density) in the first part of their trajectory. This applies to the beams accelerated between the heights of 10.5 and 12.5 Mm in Fig. \ref{fig:xz_power_change_selected_beams_characteristics_4_magmir}, as the coronal loop leg they travel along expands with depth for the first 10 Mm, before contracting again along the last stretch to the lower atmosphere. Figure \ref{fig:heating_profile_bundle_2_beam_758_magmir} shows the heating profile for one of the beams following such a trajectory.

For this beam, $Q$ remains lower throughout the corona in the model with magnetic gradient forces than in the model without. This is caused by the initial reduction in magnetic flux density with distance. From Eq. \eqref{eq:dmuds}, the negative value of $\mathrm{d}\ln B/\mathrm{d}s$ counteracts the decrease in $\mu$ due to collisions. All electrons, including the more energetic ones, thus retain a higher value of $\mu$, causing them to lose energy more slowly with distance. Since the more energetic electrons join the lower-energy part of the distribution at a reduced pace, the population at low energies decreases, reducing the energy deposition rate $Q$. This is precisely the opposite of what happens in a converging magnetic field.

\begin{figure}[!thb]
    \includegraphics{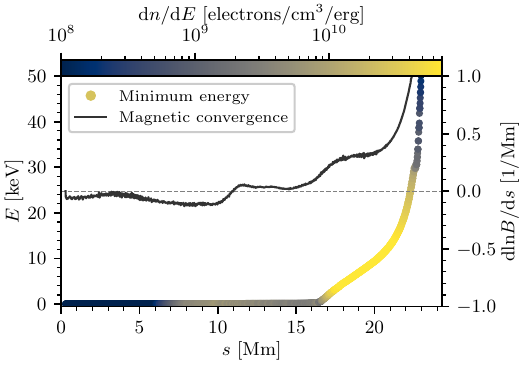}
    \centering
    \caption{Minimum energy $E$ of unreflected electrons ($\mu > 0$) remaining in the distribution as a function of distance $s$ for the beam in Fig. \ref{fig:heating_profile_bundle_2_beam_758_magmir}. The colours of the points indicate the number density $\mathrm{d}n/\mathrm{d}E$ of electrons with that energy. Included as a dark grey curve is the magnetic convergence $\mathrm{d}\ln B/\mathrm{d}s$.}
    \label{fig:mirrored_bundle_2_beam_758}
\end{figure}

Although $Q$ does exhibit a second coronal peak in the model with magnetic gradient forces due to the eventual convergence of the magnetic field, similarly to the beam in Fig. \ref{fig:heating_profile_bundle_1_beam_6265_magmir}, the energy deposition is nearly always lower than in the basic model. The only exception is at the end of the peak in the TR, as the energetic electrons that reach this depth have been subjected to slightly fewer collisions than in the basic model due to the earlier diverging magnetic field. Despite this, a significant proportion of the injected energy is clearly never deposited for the beam affected by magnetic gradient forces; upon inspection, around 50\%. The culprit is heavy magnetic mirroring in the last 7 Mm of the trajectory, where the magnetic convergence is strong. This is clear from Fig. \ref{fig:mirrored_bundle_2_beam_758}, which shows the minimum energy of electrons remaining in the distribution as a function of distance. When the magnetic convergence is negative or slightly positive, the minimum energy is close to zero, meaning that electrons are being thermalised. After $s = 16\;\mathrm{Mm}$, when the magnetic convergence increases significantly, the minimum energy increases as the electrons reach $\mu = 0$ and get reflected before they lose their energy to collisions. The energy of reflected electrons increases more and more rapidly with distance. The figure also shows with colour the number density $\mathrm{d}n/\mathrm{d}E$ of electrons with the minimum energy, indicating the number of electrons reflected at each distance. It peaks shortly after the onset of mirroring when the abundant electrons with energies around $4\;\mathrm{keV}$ are reflected. The number of reflected electrons then decreases rapidly with distance as the distribution gets depleted of increasingly energetic electrons.

The main reason this beam is particularly strongly affected by magnetic mirroring is the initial diverging magnetic field. Collisions, being more frequent at lower energies, remove less energetic electrons more efficiently and thus flatten the distribution as it propagates. By increasing $\mu$ and thus decreasing the number of collisions per unit of distance, the diverging magnetic field enables the distribution to retain not only more of its total flux but also more of its original steepness. This means that more beam flux is concentrated in the least energetic electrons when magnetic mirroring kicks in, and these are the electrons that get reflected first. The result is that more of the energy injected into the beam is lost to reflected electrons compared to what would have been lost in the absence of a diverging magnetic field.

The other notable group of beams subjected to a diverging magnetic field in set 2 are the ones accelerated near the height of 2.5 Mm around $x = 14$ to $17\;\mathrm{Mm}$ in Fig. \ref{fig:xz_power_change_selected_beams_characteristics_4_magmir} and ejected upwards along magnetic field lines that diverge with height. Some travel the length of the periodic simulation domain along the near horizontal magnetic field at the top of the domain before re-entering a coronal loop in the same downward trajectory as the beam in Fig. \ref{fig:heating_profile_bundle_2_beam_758_magmir}. They produce tiny $Q$ values in their long journey through the upper corona, particularly for the model accounting for magnetic gradient forces. Otherwise, their heating profiles exhibit the same features as seen in Fig. \ref{fig:heating_profile_bundle_2_beam_758_magmir}, and they typically lose around 25\% of their injected energy to magnetic mirroring.

\subsubsection{Current sheet leg (set 3)}

\begin{figure}[!thb]
    \includegraphics{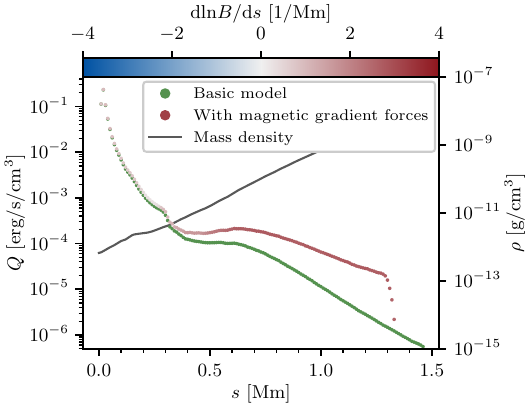}
    \centering
    \caption{Like Fig. \ref{fig:heating_profile_bundle_1_beam_6265_magmir}, but for an electron beam in set 3 in Fig. \ref{fig:xz_power_change_selected_beams_characteristics_4_magmir} that is ejected straight down into the transition region. The beam has $\mathcal{F}_{\mathrm{beam},0} \approx 6.4 \cdot 10^5\;\mathrm{erg}/\mathrm{s}/\mathrm{cm}^2$ and $E_\mathrm{c} \approx 1.8\;\mathrm{keV}$.}
    \label{fig:heating_profile_bundle_3_beam_1063_magmir}
\end{figure}

The beams in set 3 are very different from the ones in the other two sets in that they originate in a relatively dense environment and are ejected almost directly into the TR from the acceleration region. Figure \ref{fig:heating_profile_bundle_3_beam_1063_magmir} shows a typical heating profile from set 3. In both models, the vast majority of energy deposition occurs near the beginning of the trajectory owing to the immediately high density. In the basic model, $Q$ decreases monotonically with distance after the first 50 km. However, in the model with magnetic gradient forces, the decrease in $Q$ is strongly counteracted once the magnetic convergence becomes sufficiently high. The mechanism is the same as the one responsible for the second coronal peak in $Q$ for the beams starting higher in the corona: a faster decrease in $\mu$ with distance for all electrons, leading to a slight steepening of the distribution and thus more efficient energy deposition. The beam is effectively depleted after about 1 Mm of travel with enhanced energy deposition. In the basic model, the beam continues with a steady decrease in $Q$.

\begin{figure}[!thb]
    \includegraphics{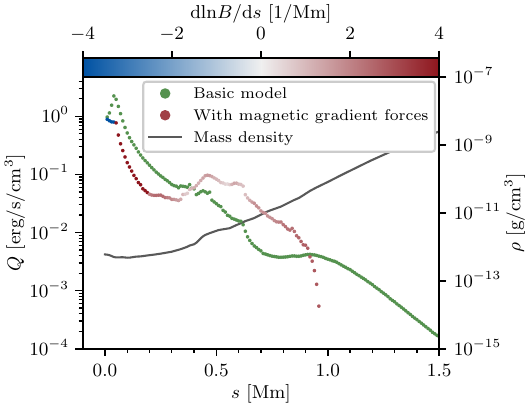}
    \centering
    \caption{Like Fig. \ref{fig:heating_profile_bundle_1_beam_6265_magmir}, but for an electron beam in set 3 in Fig. \ref{fig:xz_power_change_selected_beams_characteristics_4_magmir} whose trajectory initially is directed upwards. The beam has $\mathcal{F}_{\mathrm{beam},0} \approx 1.7 \cdot 10^7\;\mathrm{erg}/\mathrm{s}/\mathrm{cm}^2$ and $E_\mathrm{c} \approx 2.7\;\mathrm{keV}$.}
    \label{fig:heating_profile_bundle_3_beam_1539_magmir}
\end{figure}

Some of the beams in set 3 travel upwards along a diverging magnetic field for a short distance before the field line turns sharply downwards, leading the beam down into the TR. The heating profile for such a case is shown in Fig. \ref{fig:heating_profile_bundle_3_beam_1539_magmir}. In the basic model, the evolution of $Q$ with distance is effectively the same as in Fig. \ref{fig:heating_profile_bundle_3_beam_1063_magmir}. In the model with magnetic gradient forces, the initial strong decrease in magnetic flux density prevents much of the early collisional deflection of parallel velocity so that $\mu$ remains higher than in the basic model throughout the distribution. This reduces the subsequent efficiency of energy loss by collisions. The intense magnetic field convergence ensuing as soon as the beam trajectory turns downwards quickly reflects the least energetic electrons, which are relatively abundant as collisions have not significantly depleted them. Hence, the beam loses most of its electrons and about 50\% of its total energy to magnetic mirroring over the first hundreds of kilometres following the switch from magnetic field line divergence to convergence. Near $s = 400\;\mathrm{km}$, the magnetic convergence starts to diminish. This reduces the rate of magnetic reflection and allows the least energetic electrons to lose more energy to collisions before being reflected. Consequently, the deposited power $Q$ increases. After about 300 km, magnetic convergence intensifies again, and magnetic reflection rapidly depletes the remaining beam electrons.

\subsection{With non-hydrogen electrons}

We performed another run with the basic version of the CEC model configured to use an electron-to-hydrogen ratio $r_\mathrm{e}$ computed from the actual electron and hydrogen densities in the atmospheric simulation rather than using $r_\mathrm{e} = x_\mathrm{H}$. Since the electron density in the simulation includes electrons contributed from elements other than hydrogen, $r_\mathrm{e}$ exceeds the hydrogen ionisation fraction $x_\mathrm{H}$ in the parts of the atmosphere hot enough to ionise the heavier elements. In the hot corona, we have $x_\mathrm{H} = 1$ and $r_\mathrm{e} \approx 1.1$. As a result, the beams in this run typically showed a slightly elevated and earlier peak in energy deposition in the corona compared to the beams in the analytical model (where $r_\mathrm{e} = x_\mathrm{H}$). For example, the coronal peak in $Q$ for the beam in Fig. \ref{fig:heating_profile_bundle_1_beam_6265_magmir} was elevated by 5\% and occurred 1 Mm earlier.

\subsection{With helium collisions}

We performed a run to investigate the impact of including collisions with ambient helium. The only difference between the parameters of this run and the run with the most basic CEC model was that we calculated the helium-to-hydrogen ratio $r_\mathrm{He} \approx 0.085$ from the hydrogen and helium mass fractions rather than setting it to zero. The low value of $r_\mathrm{He}$ suggests that the influence of helium collisions is minor compared to the influence of collisions with electrons and hydrogen. Our results confirm this, showing only a minor increase in peak coronal energy deposition due to the extra collisions with fully ionised helium and a tiny decrease in chromospheric penetration depth caused by the extra collisions with singly ionised and neutral helium.

\subsection{With non-zero ambient temperature}

To explore the effect of accounting for the non-zero temperature of the ambient plasma, we performed another run with the basic CEC model configured to employ the warm-target transport equations (evaluating $W_E$, $W_\mu$, and $W_F$ from Eqs. \eqref{eq:warm-target-contribution-e}, \eqref{eq:warm-target-contribution-mu}, and \eqref{eq:warm-target-contribution-f}) instead of the cold-target equations. The most affected beams were those with a high temperature in the coronal part of their trajectory, like the beam in Fig. \ref{fig:heating_profile_bundle_1_beam_6265_magmir}. They experienced a modest decrease in $Q$ throughout the corona (no more than 10\%, typically less) . This is caused by the suppressed energy loss rate and slower decrease in $\mu$ for beam electrons with energies approaching the mean thermal energy, as explained by the curves in Fig. \ref{fig:warm_target_contribution_functions}.

\subsection{With the ambient electric field}

To investigate the influence of the parallel component of the ambient electric field, we performed a run with $\mathcal{E} = \mathcal{E}_\mathrm{fluid}$, computed using the electric field $\mathbf{E}_\mathrm{fluid}$ supplied by the MHD simulation. Otherwise, the model was configured like the most basic version of the CEC model. The results showed that most beams have two sections of notable parallel electric field along their trajectory; at the very beginning and at the end.

The parallel electric field at the beginning is caused by the same reconnection event that produced the beam. Depending on the details of the reconnection event and the direction of the beam, the resulting value of $\mathcal{E}_\mathrm{fluid}$ can be either positive or negative, with a magnitude typically between $10^{-9}$ and $10^{-8}$ statV/cm. The distance after which $\mathcal{E}_\mathrm{fluid}$ vanishes corresponds roughly to the extent of the corresponding blue acceleration region in Fig. \ref{fig:xz_power_change_beams_characteristics_4_magmir}, but naturally depends on how deep within the region the beam originates. For some beams, the parallel electric field near their origin is sufficiently strong to accelerate or decelerate the non-thermal electrons slightly. The result is a minor shift in energy deposition towards smaller (if decelerated) or larger (if accelerated) distances. This shift is typically no longer than 1 Mm.

The parallel electric field often present at the end of a beam trajectory is also associated with reconnection, this time in the TR and chromosphere. Here, the resistivity is generally higher than in the corona, so reconnection is more pervasive. However, the resulting electric fields are typically not strong enough to significantly influence the beam evolution, which is heavily dominated by the local high collision rates.

\subsection{With the return current}

We also performed a run including only the effects of the return current on top of the most basic model. We thus used $\mathcal{E} = \mathcal{E}_\mathrm{beam}$, with $\mathcal{E}_\mathrm{beam}$ computed from Eq. \eqref{eq:return-current-electric-field}, and included the return current heating term $Q_\mathrm{r}$ from Eq. \eqref{eq:return-current-resistive-heating} in the total heating power $Q$. The results were practically identical to those obtained without the return current. In the corona, $\mathcal{E}_\mathrm{beam}$ is of order $10^{-14}$ statV/cm for most beams. As the beams enter the TR, where the resistivity increases enormously from the extremely low coronal values, $\mathcal{E}_\mathrm{beam}$ increases by roughly three orders of magnitude. The value decreases rapidly with depth as the beams lose flux to collisions, except for the beams accelerated just above the TR (like the beams in set 3), which retain their flux for longer. Even for the most affected beams, the return current electric field is too weak to alter the flux spectrum in any way, and the return current resistive heating $Q_\mathrm{r}$ is lower than the collisional heating $Q_\mathrm{beam}$ by at least five orders of magnitude.

\section{Discussion and conclusions}
\label{sec:discussion}

Magnetic gradient forces are the most consequential phenomenon ignored in the analytical model of \citetalias{frognerAcceleratedParticleBeams2020} that can be accounted for in the CEC model. For beams travelling along coronal loops, the converging magnetic field in the lower part of the loop legs deflects the electrons, causing a peak in energy deposition followed by a substantial dip as electrons of increasingly high energy get reflected and leave the beam. When the beam enters the lower atmosphere, the resulting peak in energy deposition occurs slightly deeper since magnetic mirroring has filtered out all but the relatively energetic electrons. After this peak, energy deposition drops immediately to zero as the last electrons are reflected. For beams ejected directly into the TR from the acceleration region, the heating peaks immediately due to the high density and then decreases rapidly, but magnetic convergence can hamper this decrease for a distance until all electrons have been reflected. A diverging magnetic field in the initial part of the trajectory focuses the beam. This makes it less susceptible to collisions, reducing the collisional flattening of the distribution and immediate energy deposition. When the beam enters a converging magnetic field with a steeper distribution, more electrons and energy are lost to magnetic mirroring.

An increase in coronal energy deposition combined with reduced penetration depth in the lower atmosphere, as we see in our heating profiles with magnetic convergence, was reported by \citet{chandrashekarCollisionalHeatingNonthermal1986} (and later \citet{emslieDiagnosticsElectronheatedSolar1992}), who modified the analytical expression for $Q$ derived in \citet{emslieCollisionalInteractionBeam1978} to account for a converging magnetic field with $\mathrm{d}\ln B/\mathrm{d}s \propto n_\mathrm{H}$. In their case, however, the increase in energy deposition was more evenly distributed. It did not include the second coronal peak and associated dip evident in many of our heating profiles. This discrepancy is because $\mathrm{d}\ln B/\mathrm{d}s$ typically increases significantly more steeply with depth than $n_\mathrm{H}$ in the lower parts of the coronal loops in our atmospheric simulation. Magnetic gradient forces can thus assert their influence on the beam quite abruptly before it becomes heavily depleted by collisions, which is the cause of the second coronal peak.

Numerical simulations of electron beam transport in isolated coronal loops with varying magnetic field strength \citep[e.g.][]{leachImpulsivePhaseSolar1981, emslieDiagnosticsElectronheatedSolar1992, allredUnifiedComputationalModel2015} typically assume a purely converging magnetic field along the beam trajectory. In our atmospheric simulation, however, beams commonly find themselves in a diverging magnetic field. Our results (specifically Sect. \ref{sec:converging-bundles}) indicate that even a weakly diverging magnetic field in the initial parts of the trajectory can focus the distribution enough to resist collisional flattening and make it more susceptible to magnetic mirroring when the field subsequently converges. This suggests that magnetic mirroring may be of greater importance for collective non-thermal energy transport in realistic 3D atmospheres than one would infer from most existing simulations based on isolated loops.

Although the CEC model does not account for the reflected electrons, we can reason how they would affect the energy deposition. For the beam in Fig. \ref{fig:heating_profile_bundle_2_beam_758_magmir} (from set 2), which is particularly strongly affected by magnetic mirroring, about 45\% of the total injected energy is deposited in the corona, 5\% is deposited in the lower atmosphere, and 50\% is lost to magnetic mirroring. Figure \ref{fig:mirrored_bundle_2_beam_758} suggests that most of the mirrored energy is contained in electrons with relatively low energies reflected several megametres above the TR. Most of the energy lost to magnetic mirroring would thus be deposited in the corona. Consequently, this beam's total coronal energy deposition would roughly double if reflected electrons were accounted for. A similar analysis for the beam in Fig. \ref{fig:heating_profile_bundle_1_beam_6265_magmir} (from set 1), where magnetic mirroring is less prominent, suggests a more modest increase of about 5\% in coronal energy deposition due to reflected electrons. For beams ejected directly into the TR from the acceleration region (like the beams in set 3), the electrons are reflected in a relatively dense environment, and most of the reflected energy would likely end up close to the acceleration region. Overall, the total fraction of injected energy lost to magnetic mirroring in our simulations is not very high: about 15\% in beam set 1, 5\% in set 2, 10\% in set 3, and 5\% in the atmosphere at large.

Likely, the most prominent changes the inclusion of magnetic gradient forces would make to synthetic observables from the electron beam simulations (like those presented in \citet{bakkeAcceleratedParticleBeams2023}) would be in the thermal and non-thermal emission produced at the locations of peak energy deposition in the TR and chromosphere. While the peak deposited energy is higher without magnetic convergence, the peak typically occurs a couple hundred kilometres deeper when magnetic convergence influences the beam. This extra depth could easily mean an order of magnitude higher density and a significantly lower ambient temperature, which would affect both the observed intensity and spectral signature.

The other physical effects supported by the CEC model that were neglected in the analytical model all had a much lower impact on the energy deposition than magnetic gradient forces. Coronal energy deposition was slightly increased by including electrons from ionised elements heavier than hydrogen or collisions with ambient helium and modestly reduced by accounting for a non-zero ambient temperature. Incorporating the parallel ambient electric field shifted energy deposition slightly towards smaller or larger distances for some beams. Modelling resistive heating from the return current and energy loss to the electric field driving it had no visible effect on the results.

The insignificance of the return current in our results is not surprising. As we argued in \citetalias{frognerAcceleratedParticleBeams2020}, the fluxes of the non-thermal electron beams accelerated in our atmospheric simulation are too small to produce significant charge separation and the associated return current. Nevertheless, being able to account for this effect opens up for applying the model much more energetic beams where the return current plays an important role. To assess when the return current becomes significant, we artificially increased the energy flux $\mathcal{F}_{\mathrm{beam},0}$ injected into the beam in Fig. \ref{fig:heating_profile_bundle_1_beam_6265_magmir}. For this beam, $\mathcal{F}_{\mathrm{beam},0}$ is originally of order $10^{12}\;\mathrm{erg}/\mathrm{s}/\mathrm{cm}^2$. After increasing this flux by five orders of magnitude, the influence of the return current started to become visible as an increase in $Q$ along the first stretch of the trajectory and a decrease along the remainder.

From the limited difference using the warm-target transport equations made to our results, it is clear that the relatively high coronal temperatures of up to 2 million kelvin present in our atmospheric simulation are not sufficient to considerably influence the mean rates of change in energy and pitch angle for the non-thermal electrons. Even when we artificially increased the temperature along the coronal part of the trajectory of the beam in Fig. \ref{fig:heating_profile_bundle_1_beam_6265_magmir} from 2 to 10 million kelvin, the resulting reduction in coronal energy deposition remained minor. While the potential importance of the ambient temperature for non-thermal electron transport has been clearly demonstrated \citep{gallowayFastElectronSlowingdown2005, jeffreySpatialSpectralPolarization2014, kontarCollisionalRelaxationElectrons2015}, our results attest that this mainly comes into play in more accurate transport models where collisional diffusion of energy and pitch angle is included.

The primary weakness of the non-thermal electron transport model presented here is that it neglects velocity randomisation by setting the collisional diffusion coefficients $C_{E^2}$ and $C_{\mu^2}$ in Eq. \eqref{eq:final-fokker-planck} to zero. While the randomisation of energy only affects the evolution of the non-thermal distribution for energies approaching the mean thermal energy (see Eq. \eqref{eq:energy-diffusion-coef}), pitch angle diffusion becomes significant as soon as $|\mu|$ deviates appreciably from unity (see Eq. \eqref{eq:pitch-angle-diffusion-coef}, as well as \citet{bianRoleDiffusionTransport2016}). \citet{emslieEnergyDepositionEnergetic2018} show that the most prominent change to the energy deposition caused by accounting for pitch angle diffusion is the disappearance of the peak associated with the almost simultaneous thermalisation of electrons with energies near the lower cut-off energy $E_\mathrm{c}$ (visible, for example, at distance (b) in Fig. \ref{fig:heating_profile_bundle_1_beam_6265_magmir}). Instead, these beam electrons scatter to different pitch angles, many being prevented from propagating far due to low values of $\mu$. Hence, the highest energy deposition becomes concentrated near the point of injection. In addition, it seems plausible that by spreading out the electrons with a given energy across a range of different pitch angles, pitch angle diffusion would somewhat dampen the features associated with magnetic gradient forces apparent in the heating profiles presented here.

For the results presented in this paper, we used a power-law index of $\delta = 4$. This gives a relatively flat injected distribution that is probably not representative of the weak beam type present in our simulation \citep{linSolarHardXRay2001, kruckerHardXrayMicroflares2002}. We did this to make the beams somewhat more resistant to collisions, making it easier to isolate and present the non-collisional influences on beam transport emphasised in this paper. This also makes the presented results more relevant for understanding more energetic beams. That being said, we also performed runs with higher values of $\delta$. With the exception that a higher proportion of the injected energy is deposited in the corona for higher $\delta$ (\citetalias{frognerAcceleratedParticleBeams2020} covers the effect of varying $\delta$ in the analytical model), the results were qualitatively similar to those obtained with $\delta = 4$.

Unlike the analytical transport model used in \citetalias{frognerAcceleratedParticleBeams2020}, we did not embed the CEC model directly into the Bifrost code. Instead, we applied the model on snapshots of the simulated atmosphere outputted from Bifrost, using the standalone Rust-based tool Backstaff\footnote{\href{https://github.com/lars-frogner/Backstaff}{https://github.com/lars-frogner/Backstaff}}. As discussed in \citet{frognerImplementingAcceleratedParticle2022}, there are significant challenges in efficiently implementing global energy transport in the domain decomposition-based parallelisation scheme used in Bifrost. These difficulties are magnified with the CEC model due to the considerable amount of data associated with each beam that must be communicated between processes. While the data usage of the model could likely be optimised, this is outside the scope of this paper.

Because the electron transport simulations are run after the fact, we are currently unable to model the magnetohydrodynamic response of the ambient plasma to the non-thermal energy transport computed with the CEC model. However, our results in this paper show that the analytical transport model running inside Bifrost does a decent job of matching the energy deposition computed with the CEC model, with the only significant deviations coming from the absence of magnetic gradient forces in the analytical model.

The fact that the CEC model represents the electron flux spectrum explicitly opens up applications that the analytical model does not support. For example, the injected flux spectrum does not have to be a simple power-law but can take on an arbitrary distribution over energy. Hence, the model could be combined with a more sophisticated treatment of particle acceleration than the simple parametric acceleration model employed here. Another example is the computation of non-thermal bremsstrahlung spectra for comparison with observations.

The CEC model is a major step towards bridging the gap between the basic analytical beam propagation model used in \citetalias{frognerAcceleratedParticleBeams2020} and the state-of-the-art models based on the direct numerical solution of the Fokker--Planck equation. Our starting point for the CEC model is a fairly comprehensive version of the Fokker--Planck equation, including all relevant contributions from the Lorentz force as well as a detailed treatment of collisions. By neglecting diffusion in velocity space, we can convert the Fokker--Planck equation into a set of ordinary differential equations for the mean evolution of beam electrons. We can solve these efficiently enough to simulate millions of beams. This enables us to model non-thermal electrons in a realistic 3D atmosphere with their spatial distribution, energetics, and trajectories emerging from the atmospheric simulation rather than being prescribed ad hoc. Thus, we sacrifice some realism in the physical modelling of electron transport (no velocity randomisation) for increased realism in the context in which this modelling is applied. The trajectories that beams follow in our simulation and the conditions along them are highly diverse. As we have shown, the diversity in conditions leads to equal diversity in the behaviour of the electron beams. This sometimes exposes interesting phenomena -- like the amplification of magnetic mirroring by a preceding magnetic field line divergence -- that are easily overlooked in electron beam modelling with a more manually prescribed atmospheric structure along the trajectory.

\begin{acknowledgements}
    We thank Dr. Gordon Emslie for providing helpful clarifications on mean scattering theory. This research was supported by the Research Council of Norway through its Centres of Excellence scheme, project number 262622, and through grants of computing time from the Programme for Supercomputing.
\end{acknowledgements}

\bibliographystyle{aa}
\bibliography{references.bib}

\begin{appendix}
\onecolumn

\section{Specialising the Fokker--Planck equation}
\label{sec:appendix-fokker-planck-equation}

Let the electron distribution $f(\mathbf{r}, \mathbf{v}, t)$ be defined such that $f(\mathbf{r}, \mathbf{v}, t)\;\mathrm{d}^3 r\;\mathrm{d}^3 v$ is the number of electrons within the volume element $\mathrm{d}^3 r$ centred on position $\mathbf{r}$ with velocities within $\mathrm{d}^3 v$ of velocity $\mathbf{v}$ at time $t$. The evolution of $f(\mathbf{r}, \mathbf{v}, t)$ is governed by the Fokker--Planck equation,
\begin{equation}
    \label{eq:general-fokker-planck}
    \frac{\partial f(\mathbf{r}, \mathbf{v}, t)}{\partial t} + \mathbf{v} \cdot \nabla f(\mathbf{r}, \mathbf{v}, t) + \left(\frac{\mathrm{d}\mathbf{v}}{\mathrm{d}t}\right)_\mathrm{!C} \cdot \nabla_v f(\mathbf{r}, \mathbf{v}, t) = \left(\frac{\partial f(\mathbf{r}, \mathbf{v}, t)}{\partial t}\right)_\mathrm{coll},
\end{equation}
where $(\mathrm{d}\mathbf{v}/\mathrm{d}t)_\mathrm{!C}$ is the acceleration of the electrons due to non-collisional forces, and $(\partial f/\partial t)_\mathrm{coll}$ is the rate of change in the electron distribution due to collisions with ambient particles. We assume that the distribution reaches a steady state quickly compared to the rate of change in the background plasma so that we can ignore the explicit time dependence of the distribution:
\begin{equation}
    \frac{\partial f(\mathbf{r}, \mathbf{v}, t)}{\partial t} = 0.
\end{equation}
Since the electrons follow gyrating trajectories along a magnetic field line, it is helpful to express the position $\mathbf{r}$ in terms of the coordinates $(s, r_\perp, \phi)$, where $(r_\perp, \phi)$ are the polar coordinates of the electron within the plane perpendicular to the magnetic field vector $\mathbf{B}$ at a distance $s$ along the field line. We then have $\mathrm{d}^3 r = r_\perp\;\mathrm{d}\phi\;\mathrm{d}r_\perp\;\mathrm{d}s = \;\mathrm{d}A\;\mathrm{d}s$. In the case of gyromotion, it is reasonable to assume azimuthal symmetry, implying that no quantities in Eq. \eqref{eq:general-fokker-planck} depend on $\phi$. If we furthermore assume that the gyrating trajectories are relatively tight so that $r_\perp$ remains very small compared to the scale of distances $s$, we can ignore the dependence of the distribution on $r_\perp$. We can then consider $f(s, \mathbf{v})$ constant over its local cross-sectional area $A(s)$ and write the second term in Eq. \eqref{eq:general-fokker-planck} as
\begin{equation}
    \label{eq:position-term-s}
    \mathbf{v} \cdot \nabla f(\mathbf{r}, \mathbf{v}) = \frac{\mathrm{d}s}{\mathrm{d}t}\frac{\partial f(s, \mathbf{v})}{\partial s}.
\end{equation}
The right-hand side in the Fokker--Planck equation is given by (omitting for brevity the dependency of $f$ on $s$)
\begin{equation}
    \label{eq:general-fokker-planck-coll}
    \left(\frac{\partial f(\mathbf{v})}{\partial t}\right)_\mathrm{coll} = -\sum_i \frac{\partial}{\partial v_i}\left(\left\langle\Delta v_i\right\rangle f(\mathbf{v})\right) + \frac{1}{2}\sum_{ij} \frac{\partial^2}{\partial v_i\partial v_j}\left(\left\langle\Delta v_i\Delta v_j\right\rangle f(\mathbf{v})\right),
\end{equation}
where
\begin{equation}
    \label{eq:total-mean-rates-of-vel-change}
    \left\langle\Delta v_i\right\rangle = \sum_\alpha \left\langle\Delta v_i\right\rangle _\alpha \quad \mathrm{and} \quad \left\langle\Delta v_i\Delta v_j\right\rangle = \sum_\alpha \left\langle\Delta v_i\Delta v_j\right\rangle _\alpha.
\end{equation}
Here, $\langle\chi\rangle _\alpha$ is the mean rate of a quantity $\chi$ due to collisions with ambient particles of species $\alpha$, and is calculated as
\begin{equation}
    \label{eq:mean-collisional-rate_integral}
    \left\langle\chi\right\rangle _\alpha(\mathbf{v}) = \int_{v_\alpha} \int_\Omega \chi \left\lVert\mathbf{v} - \mathbf{v}_\alpha\right\rVert f_\alpha(\mathbf{v}_\alpha) \left(\frac{\mathrm{d}\sigma}{\mathrm{d}\Omega}\right)_\alpha\left(\mathbf{v}, \mathbf{v}_\alpha, \Omega\right)\;\mathrm{d}\Omega\;\mathrm{d}^3 v_\alpha,
\end{equation}
where $(\mathrm{d}\sigma/\mathrm{d}\Omega)_\alpha$ is the differential scattering cross-section for a collision between a beam particle and an ambient particle of species $\alpha$ with velocity $\mathbf{v}_\alpha$, and $f_\alpha$ is the distribution of the species $\alpha$ ambient particles. In Eq. \eqref{eq:general-fokker-planck-coll}, the quantity $\Delta v_i$ for which the mean rate is computed is the $i$'th component of the change in velocity $\mathbf{v}$ due to a collision.

We can write Eq. \eqref{eq:total-mean-rates-of-vel-change} as
\begin{equation}
    \label{eq:total-mean-rates-of-vel-change-split}
    \left\langle\Delta v_i\right\rangle = \sum_\mathrm{c} \left\langle\Delta v_i\right\rangle _\mathrm{c} + \sum_\mathrm{N}\left\langle\Delta v_i\right\rangle _\mathrm{N} \quad \mathrm{and} \quad \left\langle\Delta v_i\Delta v_j\right\rangle = \sum_\mathrm{c} \left\langle\Delta v_i\Delta v_j\right\rangle _\mathrm{c} + \sum_\mathrm{N}\left\langle\Delta v_i\Delta v_j\right\rangle _\mathrm{N},
\end{equation}
where c represents charged particle species, and N represents neutral particle species. By inserting the Rutherford differential scattering cross-section into Eq. \eqref{eq:mean-collisional-rate_integral}, \citet{rosenbluthFokkerPlanckEquationInverseSquare1957} derive expressions for $\left\langle\Delta v_i\right\rangle _\mathrm{c}$ and $\left\langle\Delta v_i\Delta v_j\right\rangle _\mathrm{c}$:
\begin{align}
    \left\langle\Delta v_i\right\rangle _\mathrm{c} &= -4\pi \left(1 + \frac{m_\mathrm{e}}{m_\mathrm{c}}\right) \frac{\Gamma_\mathrm{c}}{{m_\mathrm{e}}^2} \frac{\partial \phi_\mathrm{c}(\mathbf{v})}{\partial v_i} \label{eq:mean-vel-change-comp-rate} \\
    \left\langle\Delta v_i\Delta v_j\right\rangle _\mathrm{c} &= -8\pi\frac{\Gamma_\mathrm{c}}{{m_\mathrm{e}}^2} \frac{\partial^2 \psi_\mathrm{c}(\mathbf{v})}{\partial v_i\partial v_j}, \label{eq:mean-vel-change-comp-prod-rate}
\end{align}
where $m_\mathrm{e}$ is the electron mass, $m_\mathrm{c}$ is the mass of a species c particle, and
\begin{equation}
    \Gamma_\mathrm{c} = 4\pi e^4 {z_\mathrm{c}}^2 \ln\Lambda_\mathrm{c}.
\end{equation}
Here, $e$ is the elementary charge, $z_\mathrm{c}$ is the charge number, and $\ln\Lambda_\mathrm{c}$ is the Coulomb logarithm. The functions $\phi_\mathrm{c}$ and $\psi_\mathrm{c}$, defined as
\begin{align}
    \phi_\mathrm{c}(\mathbf{v}) &= -\frac{1}{4\pi}\int_{v_\mathrm{c}} \frac{f_\mathrm{c}(\mathbf{v}_\mathrm{c})}{\left\lVert\mathbf{v} - \mathbf{v}_\mathrm{c}\right\rVert}\;\mathrm{d}^3 v_\mathrm{c} \label{eq:phi-potential} \\
    \psi_\mathrm{c}(\mathbf{v}) &= -\frac{1}{8\pi}\int_{v_\mathrm{c}} \left\lVert\mathbf{v} - \mathbf{v}_\mathrm{c}\right\rVert f_\mathrm{c}(\mathbf{v}_\mathrm{c})\;\mathrm{d}^3 v_\mathrm{c}, \label{eq:psi-potential}
\end{align}
are potentials satisfying Poisson's equations:
\begin{align}
    {\nabla_v}^2\phi_\mathrm{c}(\mathbf{v}) &= f_\mathrm{c}(\mathbf{v}) \\
    {\nabla_v}^2\psi_\mathrm{c}(\mathbf{v}) &= \phi_\mathrm{c}(\mathbf{v}).
\end{align}
Disregarding, for now, the contribution of collisions with neutral particles, we can apply Eqs. \eqref{eq:mean-vel-change-comp-rate} and \eqref{eq:mean-vel-change-comp-prod-rate} so that Eq. \eqref{eq:general-fokker-planck-coll} becomes
\begin{equation}
    \label{eq:general-fokker-planck-coll-potentials}
    \left(\frac{\partial f(\mathbf{v})}{\partial t}\right)_\mathrm{coll} = \nabla_v \cdot \left(\left(\sum_\mathrm{c} \frac{4\pi\Gamma_\mathrm{c}}{{m_\mathrm{e}}^2}\left(1 + \frac{m_\mathrm{e}}{m_\mathrm{c}}\right)\nabla_v\phi_\mathrm{c}(\mathbf{v})\right)f(\mathbf{v}) - \nabla_v \cdot \left(f(\mathbf{v})\sum_\mathrm{c} \frac{4\pi\Gamma_\mathrm{c}}{{m_\mathrm{e}}^2}\nabla_v\nabla_v\psi_\mathrm{c}(\mathbf{v})\right)\right),
\end{equation}
where $\nabla_v\nabla_v\psi_\mathrm{c}$ represents the Hessian matrix of $\psi_\mathrm{c}$. Using the identity
\begin{equation}
    \nabla_v \cdot \left(f(\mathbf{v})\nabla_v\nabla_v\psi_\mathrm{c}(\mathbf{v})\right) = f(\mathbf{v})\nabla_v\left({\nabla_v}^2\psi_\mathrm{c}(\mathbf{v})\right) + \nabla_v\nabla_v\psi_\mathrm{c}(\mathbf{v})\nabla_v f(\mathbf{v}) = f(\mathbf{v})\nabla_v\phi_\mathrm{c}(\mathbf{v}) + \nabla_v\nabla_v\psi_\mathrm{c}(\mathbf{v})\nabla_v f(\mathbf{v}),
\end{equation}
we obtain an alternative version of Eq. \eqref{eq:general-fokker-planck-coll-potentials}:
\begin{equation}
    \label{eq:general-fokker-planck-coll-potentials-simp}
    \left(\frac{\partial f(\mathbf{v})}{\partial t}\right)_\mathrm{coll} = \nabla_v \cdot \left(\left(\sum_\mathrm{c} \frac{4\pi\Gamma_\mathrm{c}}{{m_\mathrm{e}}^2} \frac{m_\mathrm{e}}{m_\mathrm{c}} \nabla_v\phi_\mathrm{c}(\mathbf{v})\right)f(\mathbf{v}) - \left(\sum_\mathrm{c} \frac{4\pi\Gamma_\mathrm{c}}{{m_\mathrm{e}}^2}\nabla_v\nabla_v\psi_\mathrm{c}(\mathbf{v})\right)\nabla_v f(\mathbf{v})\right).
\end{equation}
Defining a friction vector
\begin{equation*}
    \mathbf{F}(\mathbf{v}) = -\sum_\mathrm{c}\frac{4\pi\Gamma_\mathrm{c}}{{m_\mathrm{e}}^2}\frac{m_\mathrm{e}}{m_\mathrm{c}}\nabla_v\phi_\mathrm{c}(\mathbf{v})
\end{equation*}
and a diffusion matrix
\begin{equation*}
    D(\mathbf{v}) = -\sum_\mathrm{c}\frac{4\pi\Gamma_\mathrm{c}}{{m_\mathrm{e}}^2}\nabla_v\nabla_v\psi_\mathrm{c}(\mathbf{v}),
\end{equation*}
Equation \eqref{eq:general-fokker-planck-coll-potentials-simp} becomes
\begin{equation}
\label{eq:general-fokker-planck-coll-friction-diffusion}
    \left(\frac{\partial f(\mathbf{v})}{\partial t}\right)_\mathrm{coll} = -\nabla_v \cdot \left(\mathbf{F}(\mathbf{v})f(\mathbf{v}) - D(\mathbf{v})\nabla_v f(\mathbf{v})\right) = -\nabla_v \cdot \mathbf{S}(\mathbf{v}).
\end{equation}
The velocity $\mathbf{v}$ is conveniently expressed in spherical coordinates $(v, \beta, \phi_v)$, where $v$ is the speed, the pitch angle $\beta$ is the angle between $\mathbf{v}$ and the local magnetic field vector $\mathbf{B}$, and $\phi_v$ is the angle of $\mathbf{v}$ within the plane perpendicular to $\mathbf{B}$. The velocity volume element then becomes $\mathrm{d}^3 v = v^2\sin\beta\;\mathrm{d}\phi_v\;\mathrm{d}\beta\;\mathrm{d}v$. Again, assuming azimuthal symmetry lets us ignore any dependence on $\phi_v$ and consider the distribution $f(s, v, \beta, t)$ constant for all azimuthal velocity orientations. In terms of the remaining spherical coordinates $(v, \beta)$, the divergence of a vector field $\mathbf{V}$ is
\begin{equation}
    \nabla_v \cdot \mathbf{V} = \frac{1}{v^2}\frac{\partial}{\partial v}(v^2 V_v) + \frac{1}{v \sin\beta}\frac{\partial}{\partial \beta}(V_\beta\sin\beta),
\end{equation}
the gradient of a scalar field $f$ has components
\begin{align}
    (\nabla_v f)_v &= \frac{\partial f}{\partial v} \\
    (\nabla_v f)_\beta &= \frac{1}{v}\frac{\partial f}{\partial\beta},
\end{align}
while the Hessian matrix has components
\begin{align}
    (\nabla_v\nabla_v f)_{vv} &= \frac{\partial^2 f}{\partial v^2} \\
    (\nabla_v\nabla_v f)_{v\beta} &= (\nabla_v\nabla_v f)_{\beta v} = \frac{1}{v}\frac{\partial^2 f}{\partial v \partial \beta} - \frac{1}{v^2}\frac{\partial f}{\partial \beta} \\
    (\nabla_v\nabla_v f)_{\beta\beta} &= \frac{1}{v}\frac{\partial f}{\partial v} + \frac{1}{v^2}\frac{\partial^2 f}{\partial \beta^2}.
\end{align}
In terms of $(v, \beta)$ coordinates, $\mathrm{d}s/\mathrm{d}t = v\cos\beta$, so Eq. \eqref{eq:position-term-s} becomes
\begin{equation}
    \mathbf{v} \cdot \nabla f(\mathbf{r}, \mathbf{v}) = v\cos\beta\frac{\partial f(s, \mathbf{v})}{\partial s}.
\end{equation}
The third term in Eq. \eqref{eq:general-fokker-planck} becomes
\begin{equation}
    \label{eq:velocity_term_v_beta}
    \left(\frac{\mathrm{d}\mathbf{v}}{\mathrm{d}t}\right)_\mathrm{!C} \cdot \nabla_v f(\mathrm{v}) = \left(\frac{\mathrm{d}v}{\mathrm{d}t}\right)_\mathrm{!C}\frac{\partial f(v, \beta)}{\mathrm{d}v} + \left(\frac{\mathrm{d}\beta}{\mathrm{d}t}\right)_\mathrm{!C}\frac{\partial f(v, \beta)}{\mathrm{d}\beta},
\end{equation}
while, Eq. \eqref{eq:general-fokker-planck-coll-friction-diffusion} becomes
\begin{equation}
    \label{eq:coll-term-spherical}
    \left(\frac{\partial f(v, \beta)}{\partial t}\right)_\mathrm{coll} = -\frac{1}{v^2}\frac{\partial}{\partial v}\left(v^2 S_v(v, \beta)\right) - \frac{1}{v \sin\beta}\frac{\partial}{\partial \beta}\left(S_\beta(v, \beta)\sin\beta\right),
\end{equation}
where
\begin{align}
    S_v(v, \beta) &= F_v(v, \beta) f(v, \beta) - \left(D_{vv}(v, \beta)\frac{\partial f(v, \beta)}{\partial v} + \frac{1}{v}D_{v\beta}(v, \beta)\frac{\partial f(v, \beta)}{\partial\beta}\right) \\
    S_\beta(v, \beta) &= F_\beta(v, \beta) f(v, \beta) - \left(D_{\beta v}(v, \beta)\frac{\partial f(v, \beta)}{\partial v} + \frac{1}{v}D_{\beta\beta}(v, \beta)\frac{\partial f(v, \beta)}{\partial\beta}\right).
\end{align}
The components of the friction vector are
\begin{align}
    F_v(v, \beta) &= -\sum_\mathrm{c} \frac{4\pi\Gamma_\mathrm{c}}{{m_\mathrm{e}}^2}\frac{m_\mathrm{e}}{m_\mathrm{c}}\frac{\partial \phi_\mathrm{c}(v, \beta)}{\partial v} \\
    F_\beta(v, \beta) &= -\sum_\mathrm{c} \frac{4\pi\Gamma_\mathrm{c}}{{m_\mathrm{e}}^2}\frac{m_\mathrm{e}}{m_\mathrm{c}}\frac{1}{v}\frac{\partial \phi_\mathrm{c}(v, \beta)}{\partial\beta},
\end{align}
while the diffusion matrix has components
\begin{align}
    D_{vv}(v, \beta) &= -\sum_\mathrm{c} \frac{4\pi\Gamma_\mathrm{c}}{{m_\mathrm{e}}^2}\frac{\partial^2 \psi_\mathrm{c}(v, \beta)}{\partial v^2} \\
    D_{v\beta}(v, \beta) &= D_{\beta v}(v, \beta) = -\sum_\mathrm{c} \frac{4\pi\Gamma_\mathrm{c}}{{m_\mathrm{e}}^2}\left(\frac{1}{v}\frac{\partial^2 \psi_\mathrm{c}(v, \beta)}{\partial v \partial \beta} - \frac{1}{v^2}\frac{\partial \psi_\mathrm{c}(v, \beta)}{\partial \beta}\right) \\
    D_{\beta\beta}(v, \beta) &= -\sum_\mathrm{c} \frac{4\pi\Gamma_\mathrm{c}}{{m_\mathrm{e}}^2}\left(\frac{1}{v}\frac{\partial \psi_\mathrm{c}(v, \beta)}{\partial v} + \frac{1}{v^2}\frac{\partial^2 \psi_\mathrm{c}(v, \beta)}{\partial \beta^2}\right).
\end{align}
We now assume that the ambient charged particles follow thermal Maxwell--Boltzmann distributions:
\begin{equation}
    \label{eq:thermal-distribution}
    f_\mathrm{c}(v_\mathrm{c}) = n_\mathrm{c}\left(2\pi {v_{\mathrm{tc}}}^2\right)^{-3/2}\exp\left(-\frac{{v_\mathrm{c}}^2}{2{v_{\mathrm{tc}}}^2}\right),
\end{equation}
where $v_{\mathrm{tc}} = \sqrt{k_\mathrm{B}T_\mathrm{c}/m_\mathrm{c}}$ is the mean thermal speed, $k_\mathrm{B}$ is the Boltzmann constant, and $n_\mathrm{c}$ and $T_\mathrm{c}$ are, respectively, the number density and temperature of the ambient charged particles of species c. Evaluating Eqs. \eqref{eq:phi-potential} and \eqref{eq:psi-potential} for the thermal distribution (Eq. \eqref{eq:thermal-distribution}) yields \citep{trubnikovParticleInteractionsFully1965}:
\begin{align}
    \phi_\mathrm{c}(v) &= -\frac{n_\mathrm{c}}{4\pi v}\mathrm{erf}(u_\mathrm{c}) \\
    \psi_\mathrm{c}(v) &= -\frac{n_\mathrm{c} {v_{\mathrm{tc}}}^2}{8\pi v}\left(\left(1 + 2u^2\right)\mathrm{erf}(u_\mathrm{c}) + u_\mathrm{c}\mathrm{erf}^\prime(u_\mathrm{c})\right),
\end{align}
where
\begin{equation}
    u_\mathrm{c} = \frac{v}{\sqrt{2}v_{\mathrm{tc}}},
\end{equation}
while $\mathrm{erf}(u)$ is the error function and $\mathrm{erf}^\prime(u)$ its derivative, defined by
\begin{align}
    \mathrm{erf}(u) &= \frac{2}{\sqrt{\pi}}\int_0^u \exp\left(-x^2\right)\;\mathrm{d}x \\
    \mathrm{erf}^\prime(u) &= \frac{2}{\sqrt{\pi}}\exp\left(-u^2\right).
\end{align}
The velocity friction $F_v$, velocity diffusion $D_{vv}$, and pitch angle diffusion $D_{\beta\beta}$ then become
\begin{align}
    F_v(v) &= -\sum_\mathrm{c} \frac{\Gamma_\mathrm{c} n_\mathrm{c}}{m_\mathrm{e}m_\mathrm{c}}\frac{1}{v^2}\left(\mathrm{erf}(u_\mathrm{c}) - u_\mathrm{c}\mathrm{erf}^\prime(u_\mathrm{c})\right) = -\sum_\mathrm{c} \frac{\Gamma_\mathrm{c} n_\mathrm{c}}{m_\mathrm{e}m_\mathrm{c}{v_{\mathrm{tc}}}^2}G(u_\mathrm{c}) \label{eq:velocity-friction} \\
    D_{vv}(v) &= \sum_\mathrm{c} \frac{\Gamma_\mathrm{c}n_\mathrm{c}}{{m_\mathrm{e}}^2}\frac{1}{2v}\left(\frac{\mathrm{erf}(u_\mathrm{c})}{{u_\mathrm{c}}^2} - \frac{\mathrm{erf}^\prime(u_\mathrm{c})}{u_\mathrm{c}}\right) = \sum_\mathrm{c} \frac{\Gamma_\mathrm{c} n_\mathrm{c}}{{m_\mathrm{e}}^2}\frac{G(u_\mathrm{c})}{v} \label{eq:velocity-diffusion} \\
    D_{\beta\beta}(v) &= \sum_\mathrm{c} \frac{\Gamma_\mathrm{c} n_\mathrm{c}}{{m_\mathrm{e}}^2}\frac{1}{4v}\left(\left(2 - \frac{1}{{u_\mathrm{c}}^2}\right)\mathrm{erf}(u_\mathrm{c}) + \frac{\mathrm{erf}^\prime(u_\mathrm{c})}{u_\mathrm{c}}\right) = \sum_\mathrm{c} \frac{\Gamma_\mathrm{c} n_\mathrm{c}}{{m_\mathrm{e}}^2}\frac{\mathrm{erf}(u_\mathrm{c}) - G(u_\mathrm{c})}{2v}, \label{eq:pitch-angle-diffusion}
\end{align}
where
\begin{equation}
    G(u) = \frac{\mathrm{erf}(u) - u\mathrm{erf}^\prime(u)}{2u^2}.
\end{equation}
Since the thermal distribution is isotropic, the potentials $\phi_\mathrm{c}(v)$ and $\psi_\mathrm{c}(v)$, do not depend on $\beta$, so $F_\beta = D_{v\beta} = D_{\beta v} = 0$. When collisions with neutral particles are taken into account, Eqs. \eqref{eq:velocity-friction} and \eqref{eq:pitch-angle-diffusion} obtain some additional terms \citep{evansAtomicNucleus1955a,allredModelingTransportNonthermal2020}:
\begin{align}
    F_v(v) &= -\sum_\mathrm{c} \frac{\Gamma_\mathrm{c} n_\mathrm{c}}{m_\mathrm{e}m_\mathrm{c}{v_{\mathrm{tc}}}^2}G(u_\mathrm{c}) - \sum_\mathrm{N} \frac{\Gamma_\mathrm{N}^\prime n_\mathrm{N}}{{m_\mathrm{e}}^2v^2} \label{eq:velocity-friction-with-neutrals} \\
    D_{\beta\beta}(v) &= \sum_\mathrm{c} \frac{\Gamma_\mathrm{c} n_\mathrm{c}}{{m_\mathrm{e}}^2}\frac{\mathrm{erf}(u_\mathrm{c}) - G(u_\mathrm{c})}{2v} + \sum_\mathrm{N} \frac{\Gamma_\mathrm{N}^{\prime\prime} n_\mathrm{N}}{2{m_\mathrm{e}}^2v}, \label{eq:pitch-angle-diffusion-with-neutrals}
\end{align}
where
\begin{align}
    &\Gamma_\mathrm{N}^\prime = 4\pi e^4 Z_\mathrm{N} \ln\Lambda_\mathrm{N}^\prime, \\
    &\Gamma_\mathrm{N}^{\prime\prime} = 4\pi e^4 {Z_\mathrm{N}}^2 \ln\Lambda_\mathrm{N}^{\prime\prime}.
\end{align}
Here, $Z_\mathrm{N}$ and $n_\mathrm{N}$ are, respectively, the atomic number and number density of neutral particles of species N, while $\ln\Lambda_\mathrm{N}^\prime$ and $\ln\Lambda_\mathrm{N}^{\prime\prime}$ are effective Coulomb logarithms \citep{evansAtomicNucleus1955a, snyderMultipleScatteringFast1949}. We can now write Eq. \eqref{eq:coll-term-spherical} as
\begin{equation}
    \label{eq:coll-term-spherical-thermal}
    \left(\frac{\partial f(v, \beta)}{\partial t}\right)_\mathrm{coll} = -\frac{1}{v^2}\left(\frac{\partial}{\partial v}\left(v^2 F_v(v) f(v, \beta) - v^2 D_{vv}(v)\frac{\partial f(v, \beta)}{\partial v}\right) - D_{\beta\beta}(v)\left(\cot\beta\frac{\partial f(v, \beta)}{\partial\beta} + \frac{\partial^2 f(v, \beta)}{\partial \beta^2}\right)\right),
\end{equation}
which, inserting Eqs. \eqref{eq:velocity-diffusion}, \eqref{eq:velocity-friction-with-neutrals}, and \eqref{eq:pitch-angle-diffusion-with-neutrals}, becomes
\begin{equation}
    \label{eq:coll-term-spherical-thermal-inserted}
    \begin{split}
        \left(\frac{\partial f(v, \beta)}{\partial t}\right)_\mathrm{coll} = \frac{1}{v^2}\left(\frac{\partial}{\partial v}\left(\left(\sum_\mathrm{c} \frac{2\Gamma_\mathrm{c} n_\mathrm{c}}{m_\mathrm{e}m_\mathrm{c}} {u_\mathrm{c}}^2 G(u_\mathrm{c}) + \sum_\mathrm{N} \frac{\Gamma_\mathrm{N}^\prime n_\mathrm{N}}{{m_\mathrm{e}}^2}\right) f(v, \beta) + \sum_\mathrm{c} \frac{\Gamma_\mathrm{c} n_\mathrm{c}}{{m_\mathrm{e}}^2}v G(u_\mathrm{c})\frac{\partial f(v, \beta)}{\partial v}\right) \right. \\
        \left. + \left(\sum_\mathrm{c} \frac{\Gamma_\mathrm{c} n_\mathrm{c}}{{m_\mathrm{e}}^2}\frac{\mathrm{erf}(u_\mathrm{c}) - G(u_\mathrm{c})}{2v} + \sum_\mathrm{N} \frac{\Gamma_\mathrm{N}^{\prime\prime} n_\mathrm{N}}{2{m_\mathrm{e}}^2 v}\right)\left(\cot\beta\frac{\partial f(v, \beta)}{\partial\beta} + \frac{\partial^2 f(v, \beta)}{\partial \beta^2}\right) \vphantom{\frac{\partial}{\partial v}}\right).
    \end{split}
\end{equation}
Let us here perform a change of variables from $(v, \beta)$ to $(E, \mu)$, where $E = m_\mathrm{e}v^2/2$ is the kinetic energy and $\mu = \cos\beta$. The velocity volume element then becomes
\begin{equation}
    \label{eq:velocity-volume-element-e-mu}
    \mathrm{d}^3 v = v^2\sin(\beta)\;\mathrm{d}\phi_v\;\mathrm{d}\beta\;\mathrm{d}v = \frac{v}{m_\mathrm{e}}\;\mathrm{d}\phi_v\;\mathrm{d}\mu\;\mathrm{d}E.
\end{equation}
We will also switch from using the phase-space distribution $f(v, \beta)$ to using the field-aligned electron flux spectrum $F(E, \mu)$ for representing the beam. The field-aligned electron flux spectrum is defined such that $F(E, \mu)\;\mathrm{d}\mu\;\mathrm{dE}$ is the rate of electrons with energies within $\mathrm{d}E$ of $E$ and pitch angle cosines within $\mathrm{d}\mu$ of $\mu$ flowing in the positive magnetic field direction through a unit cross-sectional area. We can express this as
\begin{equation}
    F(E, \mu)\;\mathrm{d}\mu\;\mathrm{dE} = \int_{\phi_v} \mu v f(v, \beta)\;\mathrm{d}v^3.
\end{equation}
Using Eq. \eqref{eq:velocity-volume-element-e-mu} and performing the integration over $\phi_v$, we find the relation
\begin{equation}
    \label{eq:phase-space-distribution-vs-electron-flux-spectrum}
    f(v, \beta) = \frac{m_\mathrm{e}}{2\pi \mu v^2}F(E, \mu) = \frac{{m_\mathrm{e}}^2}{4\pi \mu E}F(E, \mu).
\end{equation}
Substituting all instances of $f(v, \beta)$ with $F(E, \mu)$ using Eq. \eqref{eq:phase-space-distribution-vs-electron-flux-spectrum}, transforming all derivatives with respect to $v$ and $\beta$ to derivatives with respect to $E$ and $\mu$ and expanding them, we can write the Fokker--Planck equation as
\begin{multline}
    \frac{\partial F(E, \mu)}{\partial s} + \left(\frac{\mathrm{d}E}{\mathrm{d}s}\right)_\mathrm{!C}\frac{\partial F(E, \mu)}{\partial E} + \left(\frac{\mathrm{d}\mu}{\mathrm{d}s}\right)_\mathrm{!C}\frac{\partial F(E, \mu)}{\partial \mu} =\\
    \left(\frac{1}{E}\left(\frac{\mathrm{d}E}{\mathrm{d}s}\right)_\mathrm{!C} + \frac{1}{\mu}\left(\frac{\mathrm{d}\mu}{\mathrm{d}s}\right)_\mathrm{!C} + c_F\right)F(E, \mu) + c_E\frac{\partial F(E, \mu)}{\partial E} + c_\mu\frac{\partial F(E, \mu)}{\partial \mu} + c_{E^2}\frac{\partial^2 F(E, \mu)}{\partial E^2} + c_{\mu^2}\frac{\partial^2 F(E, \mu)}{\partial \mu^2},
\end{multline}
where we have used
\begin{align}
    \left(\frac{\mathrm{d}E}{\mathrm{d}s}\right)_\mathrm{!C} &= \frac{1}{\mu v}\left(\frac{\mathrm{d}E}{\mathrm{d}t}\right)_\mathrm{!C} \\
    \left(\frac{\mathrm{d}\mu}{\mathrm{d}s}\right)_\mathrm{!C} &= \frac{1}{\mu v}\left(\frac{\mathrm{d}\mu}{\mathrm{d}t}\right)_\mathrm{!C}.
\end{align}
The collisional coefficients are given by
\begin{multline}
    c_F = -\frac{1}{2\mu E^2}\left(\sum_\mathrm{c}\Gamma_\mathrm{c} n_\mathrm{c}\left(\frac{m_\mathrm{e}}{m_\mathrm{c}}\mathrm{erf}(u_\mathrm{c}) + \left(1 - \frac{m_\mathrm{e}}{m_\mathrm{c}}\left(1 + {u_\mathrm{c}}^2\right)\right)u_\mathrm{c}\mathrm{erf}^\prime(u_\mathrm{c}) - 4G(u_\mathrm{c})\right) + \sum_\mathrm{N}\Gamma_\mathrm{N}^\prime n_\mathrm{N} \right. \\
    \left. - \frac{1}{2\mu^2}\left(\sum_\mathrm{c}\Gamma_\mathrm{c} n_\mathrm{c}\left(\mathrm{erf}(u_\mathrm{c}) - G(u_\mathrm{c})\right) + \sum_\mathrm{N}\Gamma_\mathrm{N}^{\prime\prime} n_\mathrm{N}\right)\right)
\end{multline}
\begin{align}
    c_E &= \frac{1}{2\mu E}\left(\sum_\mathrm{c}\Gamma_\mathrm{c} n_\mathrm{c}\left(\frac{m_\mathrm{e}}{m_\mathrm{c}}\mathrm{erf}(u_\mathrm{c}) + \left(1 - \frac{m_\mathrm{e}}{m_\mathrm{c}}\right)u_\mathrm{c}\mathrm{erf}^\prime(u_\mathrm{c}) - 4G(u_\mathrm{c})\right) + \sum_\mathrm{N}\Gamma_\mathrm{N}^\prime n_\mathrm{N}\right) \\
    c_\mu &= -\frac{1}{4\mu^2 E^2}\left(\sum_\mathrm{c}\Gamma_\mathrm{c} n_\mathrm{c}\left(\mathrm{erf}(u_\mathrm{c}) - G(u_\mathrm{c})\right) + \sum_\mathrm{N}\Gamma_\mathrm{N}^{\prime\prime} n_\mathrm{N}\right) \\
    c_{E^2} &= \frac{1}{\mu}\sum_\mathrm{c}\Gamma_\mathrm{c} n_\mathrm{c}G(u_\mathrm{c}) \label{eq:energy-diffusion-coef} \\
    c_{\mu^2} &= \frac{1 - \mu^2}{8\mu E^2}\left(\sum_\mathrm{c}\Gamma_\mathrm{c} n_\mathrm{c}\left(\mathrm{erf}(u_\mathrm{c}) - G(u_\mathrm{c})\right) + \sum_\mathrm{N}\Gamma_\mathrm{N}^{\prime\prime} n_\mathrm{N}\right). \label{eq:pitch-angle-diffusion-coef}
\end{align}
From here, we can rewrite the equation to the following form:
\begin{multline}
    \frac{\partial F(E, \mu)}{\partial s} + \left(\frac{\mathrm{d}E}{\mathrm{d}s}\right)_\mathrm{!C}\frac{\partial F(E, \mu)}{\partial E} + \left(\frac{\mathrm{d}\mu}{\mathrm{d}s}\right)_\mathrm{!C}\frac{\partial F(E, \mu)}{\partial \mu} =\\
    \left(\frac{1}{E}\left(\frac{\mathrm{d}E}{\mathrm{d}s}\right)_\mathrm{!C} + \frac{1}{\mu}\left(\frac{\mathrm{d}\mu}{\mathrm{d}s}\right)_\mathrm{!C} + C_F\right)F(E, \mu) + \frac{\partial}{\partial E}\left(C_E F(E, \mu)\right) + \frac{\partial}{\partial \mu}\left(C_\mu F(E, \mu)\right) + \frac{\partial^2}{\partial E^2}\left(C_{E^2} F(E, \mu)\right) + \frac{\partial^2}{\partial \mu^2}\left(C_{\mu^2} F(E, \mu)\right).
\end{multline}
The $C$- and $c$-coefficients are related by
\begin{align}
    C_F &= c_F - \frac{\partial C_E}{\partial E} - \frac{\partial C_\mu}{\partial \mu} - \frac{\partial^2 C_{E^2}}{\partial E^2} - \frac{\partial^2 C_{\mu^2}}{\partial \mu^2} \\
    C_E &= c_E - 2\frac{\partial C_{E^2}}{\partial E} \\
    C_\mu &= c_\mu - 2\frac{\partial C_{\mu^2}}{\partial \mu} \\
    C_{E^2} &= c_{E^2} \\
    C_{\mu^2} &= c_{\mu^2}.
\end{align}
Using this, we find
\begin{align}
    C_F &= 0 \\
    C_E &= \frac{1}{2\mu E}\left(\sum_\mathrm{c}\Gamma_\mathrm{c} n_\mathrm{c}\left(\frac{m_\mathrm{e}}{m_\mathrm{c}}\mathrm{erf}(u_\mathrm{c}) - \left(1 + \frac{m_\mathrm{e}}{m_\mathrm{c}}\right)u_\mathrm{c}\mathrm{erf}^\prime(u_\mathrm{c})\right) + \sum_\mathrm{N}\Gamma_\mathrm{N}^\prime n_\mathrm{N}\right) \\
    C_\mu &= \frac{1}{4E^2}\left(\sum_\mathrm{c}\Gamma_\mathrm{c} n_\mathrm{c}\left(\mathrm{erf}(u_\mathrm{c}) - G(u_\mathrm{c})\right) + \sum_\mathrm{N}\Gamma_\mathrm{N}^{\prime\prime} n_\mathrm{N}\right).
\end{align}

\end{appendix}

\end{document}